\documentclass[twocolumn,superscriptaddress,prx,amsmath,amstex,amssymb,citeautoscript,longbibliography]{revtex4-1}
\pdfoutput=1

\usepackage{graphicx}
\usepackage{amsmath}
\usepackage{amsfonts}
\usepackage{amsthm} 
\usepackage{amssymb}
\usepackage{bm}
\usepackage{color}
\usepackage{soul} 
\usepackage{amssymb}
\usepackage{wasysym}
\usepackage{dsfont}
\usepackage{float}

\usepackage{hyperref}
\hypersetup{
    unicode=false,          
    pdftoolbar=false,        
    pdfmenubar=true,        
    pdffitwindow=false,     
    pdfstartview={FitH},    
    pdftitle={},    
    pdfauthor={Authors},     
    pdfsubject={},   
    pdfcreator={},   
    pdfproducer={}, 
    pdfkeywords={quantum many-body scars} {constrained spin models}, 
    pdfnewwindow=true,      
    colorlinks=true,       
    linkcolor=black,          
    citecolor=blue,        
    filecolor=magenta,      
    urlcolor=blue           
}

\usepackage{graphicx}
\usepackage{verbatim}

\newcommand{\ket}[1]{\mbox{$| #1 \rangle$}}
\newcommand{\bra}[1]{\mbox{$\langle #1 |$}}

\newcommand{\HG}[2]{G_{#1}^{#2} }

\usepackage{soul}
\usepackage[normalem]{ulem}
\newtheorem*{lemma*}{Lemma}
\usepackage{algorithm}
\usepackage{algpseudocode}

\newtheorem{dfn}{Definition}

\newcommand*{\id}{{\normalfont\hbox{1\kern-0.15em \vrule width .8pt depth-.5pt}}}

\usepackage{times}

\begin{document}

\title{Hypergrid subgraphs and the origin of scarred quantum walks in many-body Hilbert space}

\author{Jean-Yves Desaules}
\affiliation{School of Physics and Astronomy, University of Leeds, Leeds LS2 9JT, United Kingdom}
\author{Kieran Bull}
\affiliation{School of Physics and Astronomy, University of Leeds, Leeds LS2 9JT, United Kingdom}
\author{Aiden Daniel}
\affiliation{School of Physics and Astronomy, University of Leeds, Leeds LS2 9JT, United Kingdom}
\author{Zlatko Papi\'c}
\affiliation{School of Physics and Astronomy, University of Leeds, Leeds LS2 9JT, United Kingdom}
\date{\today}

\begin{abstract}
Following the recent observation of wave function revivals in large Rydberg atom quantum simulators, much effort has focused on understanding the emergence of  many-body scars  in non-integrable quantum systems. Here we explore the origin of scarred  wavefunction revivals in a family of models obtained by deforming the graph adjacency matrix of the PXP model -- the effective model of Rydberg atoms in the strong Rydberg blockade regime. We consider deformations that either enhance the Rydberg constraint, ultimately resulting in an effective tight-binding model of two hypercubes joined at a single vertex, or relax the constraint until reaching the free spin-1/2 model. In the former case, we argue that the model of two joined hypercubes captures the essential features of many-body scarring present in the PXP model. On the other hand, relaxing the constraint leads to a sequence of new  scarred models, some with more robust scarring signatures than the PXP model, as can be understood from the graph-theoretic viewpoint. Our results shed light on the nature of scarring in the PXP model by identifying its simple parent model, while also highlighting its distinction from the free-spin precession.  
\end{abstract}

\maketitle{}

\section{Introduction}
\label{sec:introduction}

Quantum revival -- a phenomenon where the wave function $\ket{\psi(T)}$ at some time $t=T$ returns to its value at initial time $t=0$~\cite{Bocchieri1957,Percival1961}, i.e., 
$|\langle \psi(0) |  \psi(T) \rangle|^2 \sim O(1)$,
has played an important role in understanding coherence properties of few-body or weakly-interacting quantum systems~\cite{Eberly1980,Rempe1987,Yeazell1990,Baumert1992,Brune1996,Aronstein1997, Robinett2002,Greiner2002,Will2010,Schweigler2017,Dubois2017,Rauer2018}. 
The ability to engineer recurrent behaviour in more complex quantum systems is an important task as it allows one to study their long-term coherent evolution beyond the initial relaxation, while on the other hand, it also provides insight into the emergence of statistical ensembles in closed systems that evolve according to the Schr\"odinger unitary dynamics.

Beyond material systems, it has been fruitful to study wave function revivals in a more abstract setting of quantum walks on various types of graphs~\cite{Godsil2012}.  In physics, one of the ubiquitous graphs is the hypercube graph, which arises as  the adjacency matrix of a system of free spin-1/2 degrees of freedom, 
\begin{eqnarray}
H=\sum_{j=1}^N X_j, \label{eq:para}
\end{eqnarray}
where $X_j$ is the standard Pauli-$x$ matrix on site $j$. This graph has  $2^N$ vertices which are product states of spins, usually taken to be oriented along the $z$-axis,  $|\sigma_1, \sigma_2 \ldots \sigma_N\rangle$, with $\sigma_i{=}0,1$. Moreover, this graph is unweighted, i.e., all edges of the graph are equal to 1 because the  Hamiltonian matrix elements between different basis states are all the same. The hypercube graph is known to support perfect state transfer (and thus perfect quantum revival) from any vertex -- a fact that has attracted significant attention in both graph theory~\cite{Godsil2012} and physics community~\cite{Bose2003,Christandl2004,Kay2010} where it is used as a starting point for designing quantum network architectures, e.g., see a recent realisation in superconducting qubits in Ref.~\onlinecite{Li2018}. More generally, there has been growing interest in using graph theory to describe universal properties of quantum many-body systems~\cite{Decamp2020_1,Decamp2020_2} and their thermalization dynamics~\cite{RoyLazarides2020}.

Intriguingly, even the simple quantum systems, such as a free particle hopping on a 1D lattice, require careful tuning of the hopping amplitudes in order to sustain perfect quantum state transfer~\cite{Christandl2004}. In generic many-body systems, perfect state transfer is expected to be exceptionally rare, if not impossible, once the  wave function is allowed to spread across an exponentially large Hilbert space. Nevertheless, as recent experiments on arrays of Rydberg atoms have shown~\cite{Labuhn2016,Bernien2017, Bluvstein2021}, in certain ``quantum many-body scarred"  systems, some initial states can undergo robust state transfer, despite the fact that the system overall is non-integrable, i.e., does not have any global conserved quantities apart from total energy. This phenomenon is now understood to be a consequence of a small set of non-thermal eigenstates dispersed throughout the many-body energy spectrum~\cite{Turner2017, TurnerPRB}. These eigenstates were shown to form an approximate representation of an su(2) algebra~\cite{Choi2018}, thus they can be visualised as basis states of an emergent ``big spin". The underlying su(2) algebra gives rise to an equal energy separation between the special eigenstates, and the observed revival dynamics can be understood as  semiclassical spin precession~\cite{wenwei18TDVPscar}. This phenomenology forms the core of quantum many-body scarring -- a many-body version of the phenomena associated with a  single particle confined to a chaotic stadium billiard~\cite{Heller84}. For recent introductions to many-body quantum scars, see Refs.~\cite{Serbyn2021, PapicReview,MoudgalyaReview}.

From a graph point of view, the prototype model of quantum many-body scarring in Rydberg atom chains -- the so-called PXP model~\cite{FendleySachdev,Lesanovsky2012} -- is  a \emph{partial cube}, i.e., a hypercube with some of the vertices removed (without changing the distance between any two vertices). 
The PXP  Hamiltonian is :
\begin{eqnarray}\label{eq:p_cube}
H_\mathrm{PXP} = \mathcal{P} \left( \sum_{j=1}^N X_j \right) \mathcal{P},
\end{eqnarray}
where we assume periodic boundary conditions (PBC) as we will do in the rest of this work. $\mathcal{P} = \prod_i \left[1 - (1+ Z_i)(1 + Z_{i+1})/4\right]$ is a global projector expressed in terms of the $Z$ Pauli matrix. This projector removes vertices that contain nearest-neighbour pairs of atoms that are simultaneously excited, i.e., any spin configurations of the form $|\ldots 11 \ldots\rangle$. In analogy with the Rydberg atom case, we will use the terms `excitation' and `spin-up' interchangeably in the rest of this paper. The number of vertices corresponding to the states violating the constraint scales exponentially with system size, and their removal profoundly alters the behaviour of the model: unlike the free spin-1/2 model in Eq.~(\ref{eq:para}), which is a (non-interacting) integrable model, the PXP model in Eq.~(\ref{eq:p_cube}) is chaotic~\cite{Turner2017}. 

\begin{figure}[bt]
\centering
\includegraphics[width=\linewidth]{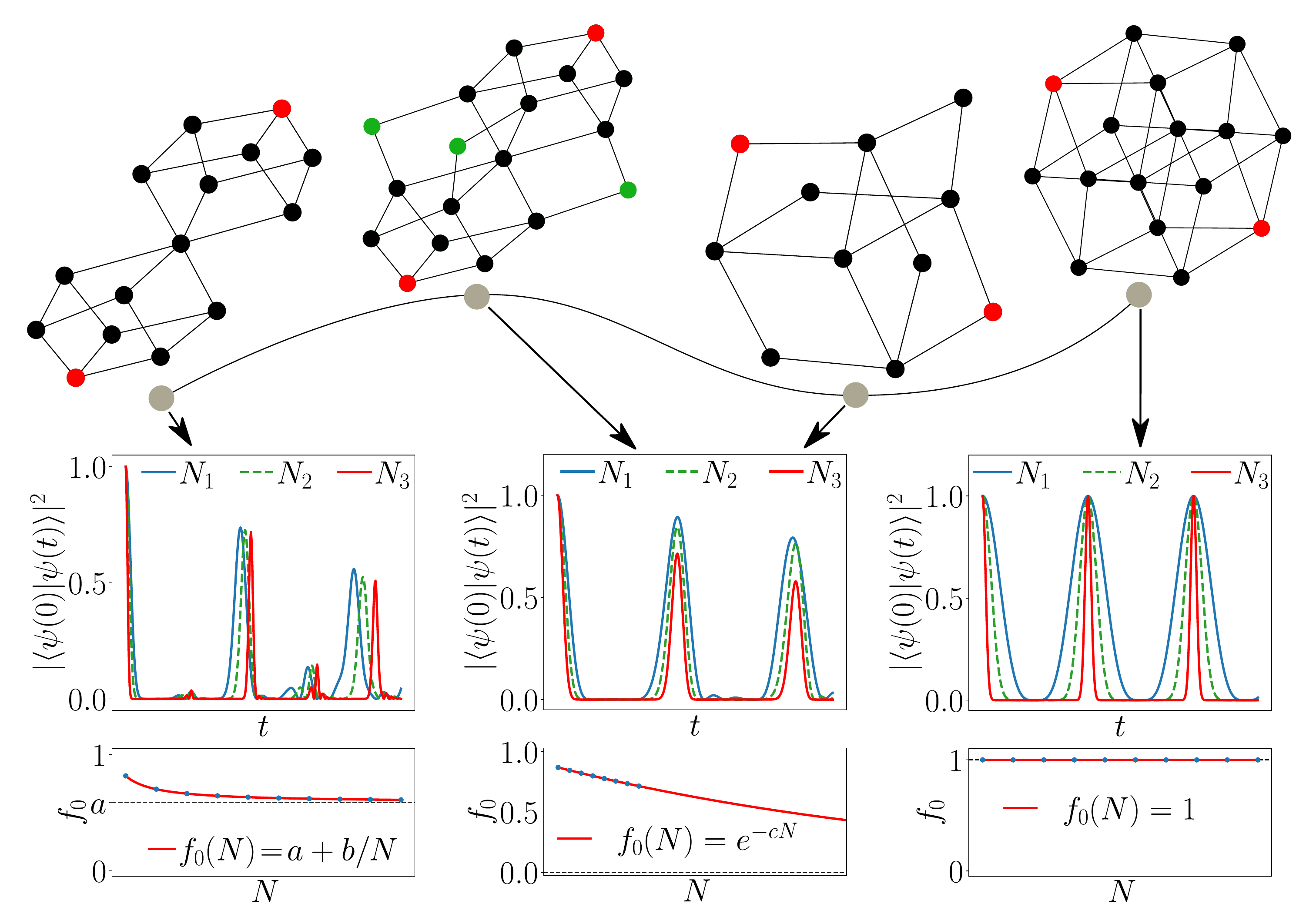}
\caption{Two families of constrained models, obtained by either strengthening or weakening the PXP constraint. The Hamiltonian adjacency graphs of the models are (from left to right): two  hypercubes joined at a single vertex, Eq.~(\ref{eq:2hc}), with $N{=}6$ spins; PXP model in Eq.~(\ref{eq:p_cube}) with $N{=}6$; a new model we refer to as ``(2,3) model" with $N{=}4$ introduced in Eq.~(\ref{eq:kmodels}) below; the free spin-1/2 model in Eq.~(\ref{eq:para}) with $N{=}4$. The red vertices denote the two N\'eel states. The black vertices in the PXP graph highlight the embedded subgraph corresponding to two hypercubes. 
The two hypercubes are connected via ``bridges" (green vertices), i.e., vertices and edges present in the PXP model but not contained within the two hypercubes.  The behaviour of the wave function fidelity revivals $|\langle \psi(0)|\psi(t) \rangle|^2$ for three different system sizes $N_1<N_2<N_3$ is sketched below the models, along with the system-size scaling of the first revival peak $f_0$.
}
\label{fig:graph_struct}
\end{figure}

Indeed, for most initial states that are product states of spins in the computational basis (i.e., vertices of the partial cube), the PXP model exhibits fast equilibration without revivals. However, for special initial states, such as the N\'eel or ``$\mathbb{Z}_2$" state, $\ket{\mathbb{Z}_2}\equiv \ket{101010\ldots}$, the PXP model undergoes a significant (albeit not perfect) state transfer to the translated N\'eel state, $\ket{010101\ldots}$. These states feature a robust quantum revival with return probability on the order $\sim 70\%$ in relatively large systems of $N{=}32$ spins~\cite{TurnerPRB} (see also Fig.~\ref{fig:graph_struct} for an example). While the existence of quantum revivals in the PXP model has been been accounted for by an emergent su(2) spectrum-generating algebra~\cite{Choi2018}, the origin of the effect remains unclear: what is it about the PXP model that allows this effective su(2) spin to arise in the first place? In particular, why are the constrained models predisposed towards this type of behaviour?

Here we explore the previous questions by taking a  graph point of  view. We demonstrate the existence of large regular subgraphs, embedded in the Hamiltonian adjacency graph,  which capture the essential physics of the revivals. In the PXP model, we will show that the relevant  subgraph consists of two large hypercubes of dimension $N/2$ each, whose corners are the two N\'eel states -- see an example in Fig.~\ref{fig:graph_struct}. The two hypercubes share a single vertex -- the polarised state $\ket{000000\ldots}$ -- and the opposite corners to this state are the two N\'eel states.  Thus, the state transfer between the two N\'eel states can be schematically understood as corner-to-corner transmission from $\ket{101010\ldots}$ to $\ket{000000\ldots}$ in the first hypercube, and then from $\ket{000000\ldots}$ to $\ket{0101010\ldots}$ in the second hypercube. Given the strong link between perfect state transfer and hypercube graphs, we  seek explanation of the revivals in the PXP model as due to the existence of two large hypercubes embedded in it, which support finite revival on their own in the thermodynamic limit, as shown below.

Specifically, our discussion of quantum revivals below will focus on three typical scenarios that are illustrated in Fig.~\ref{fig:graph_struct}. The free paramagnet in Eq.~(\ref{eq:para}) exhibits perfect revivals at arbitrarily late times and its fidelity of the first revival peak $f_0 = 1$ for any system size $N$. Other models, such as the two-hypercube model, which will be shown to have an effective single-particle description, display imperfect revivals with a certain decay time scale. Nevertheless, such models still exhibit finite revival peaks with a fixed period $T$ in the thermodynamic limit, i.e., $f_0 {\to} \mathrm{const}$ as $N {\to} \infty$. Finally, for the many-body models we consider, including the PXP model in Eq.~(\ref{eq:p_cube}), the fidelity generically decays exponentially with system size, $f_0 \propto \exp(-c N)$, thus the return probability vanishes in the thermodynamic limit. However, the fidelity \emph{density}, $\ln(f_0)/N$, for such models and special initial states considered, takes a value much closer to 0 than expected in a generic thermalising system, this signalling weak ergodicity breaking. We focus on these three classes of regular fidelity revivals with a well-defined frequency and finite period, as opposed to other types of ``accidental" revivals at irregular times or revivals due to finite size effects, for which the period increases drastically with $N$.    

In this paper we investigate two families of constrained models described by Hamiltonians of the form in Eq.~(\ref{eq:p_cube}) with different choices of projectors $\mathcal{P}$. The projector $\mathcal{P}$ is further restricted to only forbid excitations, meaning that it must always be possible to deexcite an atom. As a consequence, all graphs considered are not only partial cubes but so-called daisy cubes~\cite{Klavzar2019}. 
The first family, studied in Sec.~\ref{sec:pxp_2cube},  is obtained by strengthening the PXP constraint to turn it into the two-hypercube model. These models exhibit revivals from the N\'eel state and a band of scarred eigenstates. The latter span a subspace that evolves ``adiabatically" as the constraint is tuned between different models belonging to this family.  In contrast, the models investigated in Sec.~\ref{sec:kkp} are obtained by \emph{weakening} the constraint in a way that smoothly interpolates between the PXP model and the free spin-1/2 model. Interestingly, the many-body scarring properties of the models in this family are found to vary non-monotonically: while some models exhibit enhanced  scarring compared to the PXP model, other models ``closer" to the free spin-1/2 model actually have poorer scarring properties. This family of models thus illustrates the many-body nature of scarring in the PXP model and its lack of a simple adiabatic continuity with the free spin-1/2 model. 
Finally, in Sec.~\ref{sec:pxp_bridges} we demonstrate the robustness of our results by sampling random models with the same graph structure and showing they yield the same phenomenology. Our conclusions are presented in Sec.~\ref{sec:conc}, while Appendices contain further results including different ways of joining the hypercubes and random sampling of models. 

\section{The model of two corner-sharing hypercubes and interpolation to the PXP model}\label{sec:pxp_2cube}

A single hypercube of dimension $N$ represents a non-interacting chain of $N$ spins defined in Eq.~(\ref{eq:para}). As we explain below,  although the solution of the hypercube is well-known from the theory of angular momentum in quantum mechanics, the same problem can also be solved by mapping to a tight-binding chain with $N{+}1$ sites. The latter approach, known as the Forward Scattering Approximation (FSA), will be used throughout this paper. 
We first provide a brief overview of this formalism for a single hypercube, following Refs.~\onlinecite{Turner2017, TurnerPRB}. After this, we turn to the study of a model of two hypercubes joined at a single vertex, demonstrating that it provides a simplified description of scarred dynamics in the PXP model.

\subsection{Forward scattering approximation for the hypercube}\label{sec:fsa1cube}

The FSA method is a version of the Lanczos recurrence~\cite{lanczosbook} whereby one projects the Hamiltonian into a Krylov subspace. The usual Lanczos iteration starts with a given vector in the Hilbert space, $\ket{v_0}$, usually chosen to be random. The orthonormal basis is constructed by recursive application of the Hamiltonian $H$  to the starting vector.  The basis vector $\ket{v_{j+1}}$ is obtained from $\ket{v_j}$ by applying  $H$ and orthogonalising against $\ket{v_{j-1}}$:
\begin{equation}
 \label{eqn:lanczos}
 \beta_{j+1} \ket{v_{j+1}} = H \ket{v_{j}} - \alpha_{j} \ket{v_{j}} - \beta_{j} \ket{v_{j-1}},
\end{equation}
where $\alpha_j {=} \left<v_j|H|v_j\right>$ and $\beta{>} 0$ are chosen such that $\|v_j\| {=} 1$. Here we observe that the action of $H$ results in the next vector $\ket{v_{j+1}}$ (``forward propagation''), but also gives some weight on the previous basis vector, $\ket{v_{j-1}}$ (``backward propagation'').  

For a single hypercube in Eq.~(\ref{eq:para}), the above scheme is fully analytically tractable. Let us choose the N\'eel state $\ket{v_0} = \ket{\mathbb{Z}_2}=\ket{1010\ldots}$ as the initial vector. Moreover, we split the Hamiltonian in Eq.~(\ref{eq:para}) as $H = \sum_j X_j = H^++H^-$, i.e., into a sum of the forward and backward propagators,
\begin{subequations}\label{eqn:split}
\begin{align}\label{eqn:split1}
  H^{+} &= \sum_{j\in \text{ odd}} \sigma_j^- + \sum_{j\in \text{ even}} \sigma_j^+,
  \\ \label{eqn:split2}
  H^{-} &= \sum_{j\in \text{ odd}} \sigma_j^+ + \sum_{j\in \text{ even}} \sigma_j^-,
\end{align}
\end{subequations}
where $\sigma^\pm$ are the standard Pauli raising/lowering operators. For a single hypercube, it can be seen that $H^+$ and $H^-$ obey the standard algebra of spin raising and lowering operators. This can be used to immediately  write down the Hamiltonian matrix. Nevertheless, we will show the same result can be obtained via a sligthly different procedure, which  directly generalizes to the two-hypercube and PXP models. 

Consider the first step of the recurrence~(\ref{eqn:lanczos}). Operator $H^-$ annihilates the state $\ket{1010\ldots}$, and we obtain the vector $\beta_1 \ket{v_1} = H^+ \ket{\mathbb{Z}_2}$, which is an equal-weight superposition of all single-spin flips on top of $\ket{\mathbb{Z}_2}$, 
\begin{equation}\label{Eq:one-defect}
\beta_1  \ket{v_1} =\ket{001010\ldots}+\ket{111010\ldots}+\ket{100010\ldots}+\ldots.
\end{equation}
Hence, $H^+$ implements forward propagation while the action of $H^-$ has vanished. The vector $\ket{v_1}$ is automatically orthogonal to $\ket{v_0}$, thus we set $\alpha_0{=}0$, and $\beta_1{=}\sqrt{N}$ by normalisation, where $N$ is the number of spins.

In the second step, we observe that the action of $H^+$ on $\ket{v_1}$ will produce a state containing a pair of defects atop the N\'eel state, which is thus orthogonal to both $\ket{v_{1}}$, and $\ket{v_0}$. On the other hand, the action of the backward-scattering part gives us the original state $\ket{v_0}$, $H^- \ket{v_1} =  \beta_1 \ket{v_0}$, where we explicitly used the value of $\beta_1$. For a hypercube, one can show that
\begin{equation}\label{Eq:backward-scatter}
H^-  \ket{v_{j}} =  \beta_{j} \ket{v_{j-1}}
\end{equation}
holds more generally at \emph{every} step of the iteration. This allows to cancel $H^- \ket{v_{j}} $ with the last term in Eq.~(\ref{eqn:lanczos}), yielding the FSA recurrence:
\begin{equation}\label{Eq:lanczosFSA}
  \beta_{j+1} \ket{v_{j+1}} = H^+ \ket{v_{j}},
\end{equation}
where we also omitted the $\alpha_{j}\ket{ v_{j}}$ term since all $\alpha_j{=}0$. This follows from the fact that $H^\pm$ operators change the Hamming distance from $\ket{\mathbb{Z}_2}$ state by $\pm 1$. Hence, the new state $\ket{v_{j+1}}$ is always orthogonal to $\ket{v_{j}}$. Moreover, by the same argument, the FSA recurrence closes after $N{+}1$ steps as it reaches the vector $\ket{v_{N}} = \ket{\mathbb{Z}'_2}=\ket{0101\ldots}$, which is the translated N\'eel state that  vanishes under the action of $H^+$. 

Finally, using induction one can show  
\begin{eqnarray}\label{eq:betajmain}
  \beta_j = \sqrt{j(N - j + 1)},
\end{eqnarray}
which, as anticipated, is the well-known matrix element of a spin ladder operator.
This results in the effective tridiagonal matrix form in the basis of $\ket{v_j}$:
\begin{equation}\label{eq:HCTB}
H_\mathrm{hypercube} =\sum_{j=1}^{N} \beta_{j} \; \ket{v_j}\bra{v_{j-1}}+\mathrm{{H}.c.}
\end{equation}
This allows to reduce the dynamics from the N\'eel state to that of a tight-binding chain with the corresponding hopping strength, as shown in Fig.~\ref{fig:HC_TB}(a).
Taking into account the expression for $\beta_j$, we see that this matrix coincides with the $2 S^x$ operator for a spin of size $N/2$, resulting in a set of $N{+}1$ equidistant energy levels. Likewise, the wave functions in the basis of $\ket{v_j}$ can be obtained from the Wigner rotation matrix. 

Beyond the elementary hypercube example described above, generalisations of the FSA -- Eq.~(\ref{Eq:lanczosFSA}) -- were shown to be a useful scheme for approximating many-body scarred eigenstates in PXP and similar models~\cite{TurnerPRB, Bull2019, Bull2020,Moudgalya2019}, see Appendix~\ref{sec:fsa} for more details. For the model of two hypercubes joined at a single vertex, the FSA is also numerically exact for certain initial states. Finally, in the PXP model the FSA is no longer exact, meaning that $H^-H^+\ket{v_j}$ is not proportional to vector $\ket{v_j}$  for some values of $j$. This in turn implies that the FSA subspace is not disconnected form the rest of the Hilbert space. However, the FSA still yields highly accurate approximations of scarred eigenstates for the PXP  model~\cite{TurnerPRB} and it forms the foundation for algebraic understanding of many-body scars~\cite{Choi2018}.

\subsection{From one to two hypercubes}\label{sec:twocubes}

Now we consider in detail the problem of two hypercubes (of dimension $N/2$ each), sharing a single vertex, as in Fig.~\ref{fig:graph_struct}.
This two-hypercube model can be written as a translation-invariant spin Hamiltonian:
\begin{eqnarray}\label{eq:2hc}
\notag H_\mathrm{2HC} = \sum_j \cdots \id_{j-4} P_{j-3} \id_{j-2} P_{j-1} X_j P_{j+1}\id_{j+2}P_{j+3}\id_{j+4}\cdots,\\
\end{eqnarray}
obtained by dressing each Pauli matrix $X_j$ by an infinite string of operators alternating between identity $\id$ and local projector $P_j{=}(1-Z_j)/2$. For a finite system, the length of the string can be limited to $N/2$ on each side.   

We can express the Hamiltonian in Eq.~(\ref{eq:2hc}) in the form of Eq.~(\ref{eq:p_cube}), where global $\mathcal{P}$ now annihilates any state with excitations on both the odd and even sublattices. Thus, only one of these two sublattices can have excitations for any given state, but there is no further constraint within each sublattice. We are then left with two free spin-1/2 models with $N/2$ states (one for each sublattice) that share a single state, namely the polarised state $\ket{00\ldots 00}$.
Hence the corresponding graph consists of  two hypercubes of dimension $N/2$ sharing a single vertex. This formulation of the two-hypercube model also  makes it more transparent why it physically emerges in the PXP model (and in the free spin-1/2 model) in the first place,  as can be seen in Fig.~\ref{fig:graph_struct}.
The constraint is similar in both models, but in the case of two hypercubes it encompasses \emph{all sites} on the other sublattice instead of just the nearest neighbours in the PXP model. Thus, all states in the two-hypercube model satisfy the PXP constraint but not the other way round.

\begin{figure}[tb]
\centering
\includegraphics[width=\linewidth]{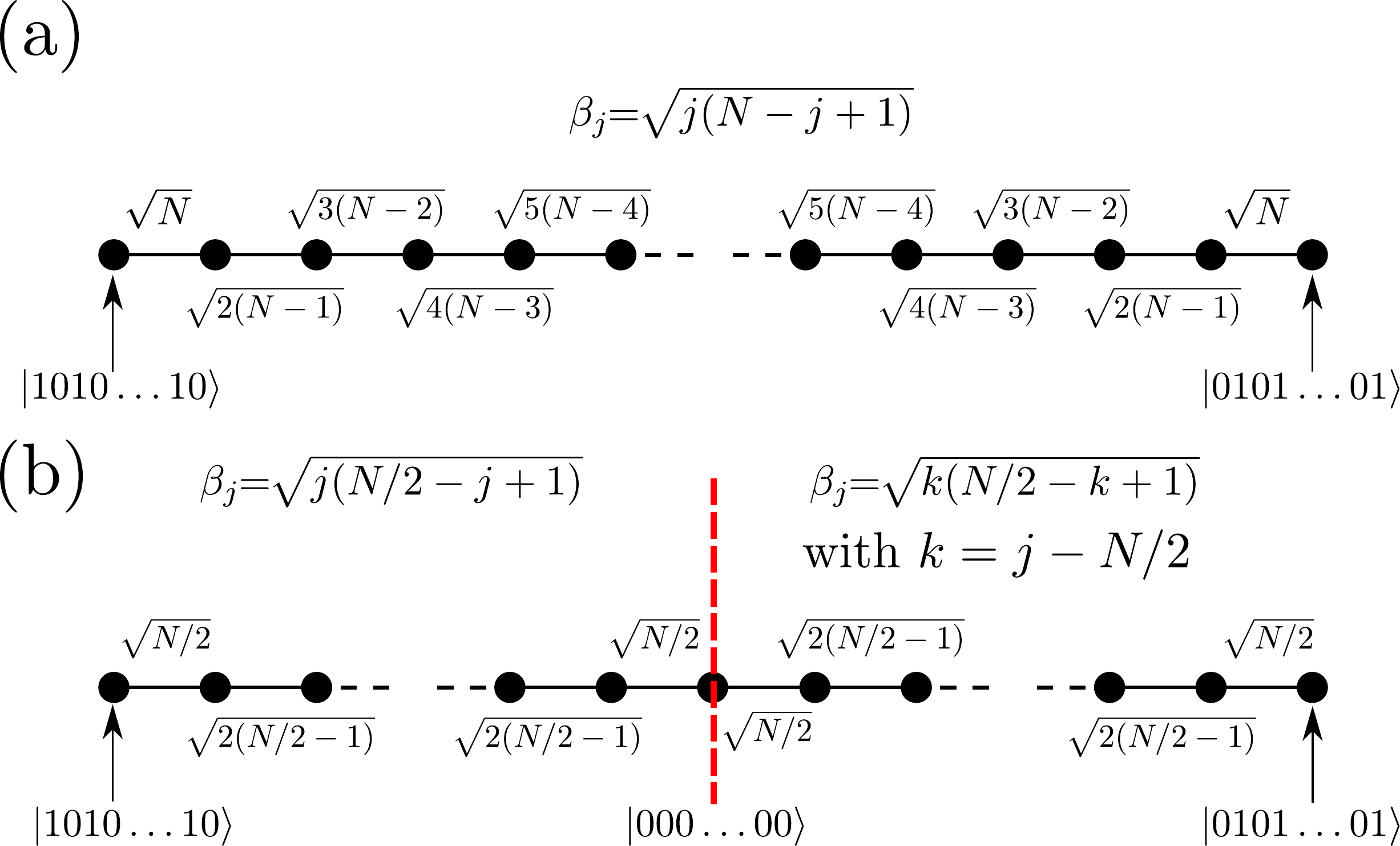}
\caption{Tight-binding chain describing the dynamics from the N\'eel initial state for (a) the free paramagnet and (b) the two-hypercube model in Eq.~(\ref{eq:2hc}). In both cases, the many-body dynamics can be reduced to a single-particle hopping on a 1D tight-binding chain, with the site-dependent hopping amplitudes indicated on the chains. From these amplitudes, it can be seen that the two-hypercube model with $N$ sites ($N$-even) simply corresponds to ``sewing" together two free paramagnets with $N/2$ sites each.}
\label{fig:HC_TB}
\end{figure}

\begin{figure}[tb]
\centering
\includegraphics[width=\linewidth]{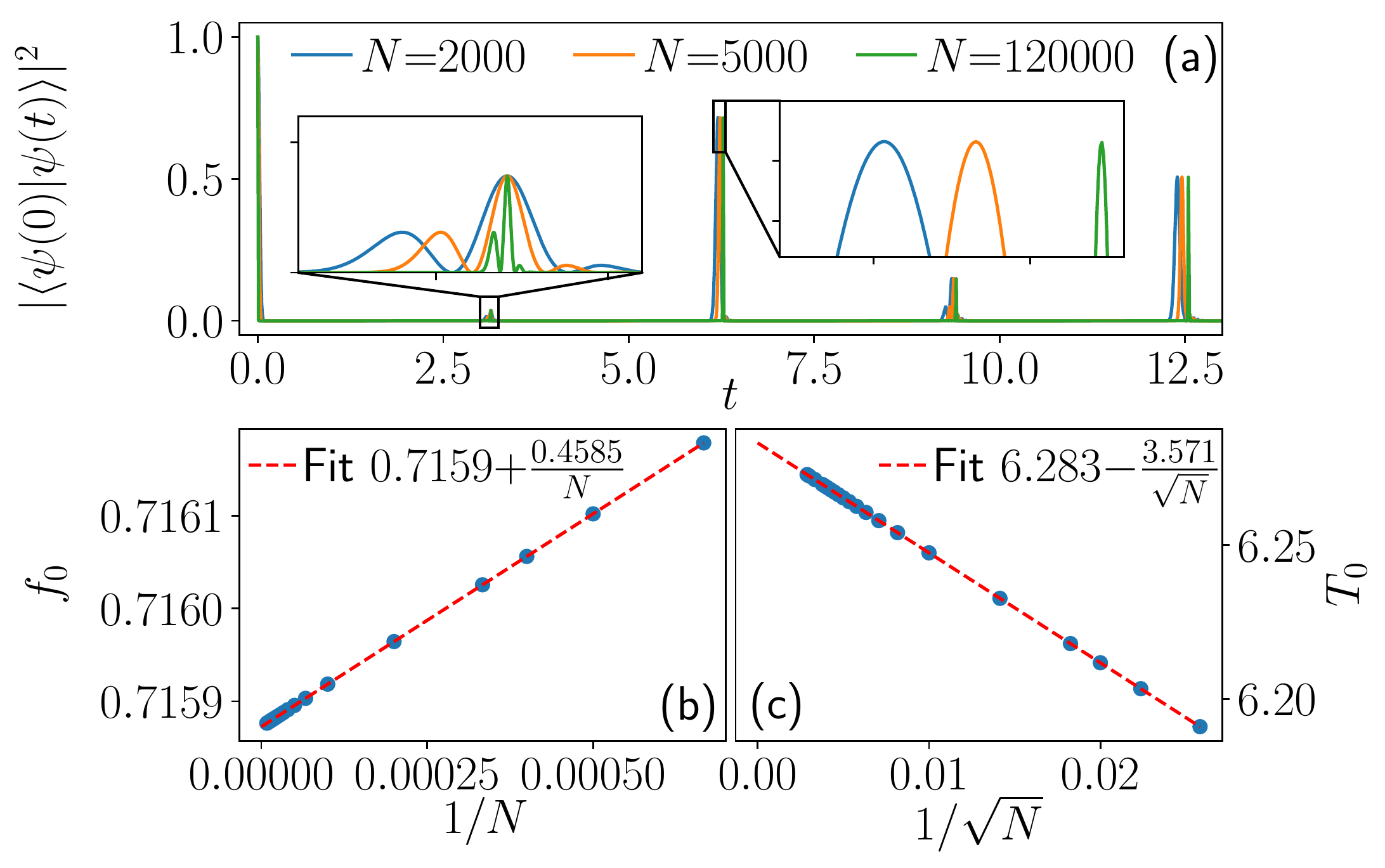}
\caption{ Dynamics in the two-hypercube model in Eq.~(\ref{eq:2hc}). (a) Time evolution of the return probability, $|\langle \psi(0)|\psi(t)\rangle|^2$,  for the  N\'eel state, $|\psi(0)\rangle{=}|1010\ldots\rangle$ and several system sizes $N$. The insets zoom in on the first revival at time $T$ and on the reflection peak at $T/2$. (b)-(c) Finite size scaling analysis of the first revival amplitude $f_0$ (b) and period $T_0$ (c). Extrapolation to $N{\to}\infty$ yields a finite revival peak $f_0$.}
\label{fig:2HC_dyn}
\end{figure}

For the two-hypercube model, the FSA introduced in Sec.~\ref{sec:fsa} remains exact for certain initial states. 
Indeed, when starting from the $|\mathbb{Z}_2\rangle=|1010\ldots\rangle$ state (which is the vertex at the maximal distance away from the shared vertex), the FSA procedure mirrors that in a single hypercube until the shared vertex is reached (after $N/2$ steps).  At that point, the second half of the FSA procedure happens exclusively in the second hypercube until the translated $|\bar{\mathbb{Z}}_2\rangle=|0101\ldots\rangle$ state is reached -- see Fig.~\ref{fig:HC_TB}(b) for an illustration. 
As a consequence, the FSA for the model in Eq.~(\ref{eq:2hc}) is exact for the $|\mathbb{Z}_2\rangle$ initial state and the tridiagonal Hamiltonian corresponds to two copies of Eq.~(\ref{eq:HCTB}) joined together.
Unfortunately, analytical diagonalisation of this Hamiltonian in the FSA subspace is no longer trivial.  Nevertheless,  due to the complexity of the problem growing only linearly with $N$, numerical  simulations on large systems $N{\lesssim} 10^5$ are possible. 
The finite-size scaling analysis in Fig.~\ref{fig:2HC_dyn} for the two-hypercube model in Eq.~(\ref{eq:2hc}) shows that this model supports revival of the wave function in the thermodynamic limit. We evaluated the quantum fidelity, $|\langle \psi(0)|e^{-i H_\mathrm{2HC} t}|\psi(0)\rangle|^2$, which is seen to rapidly decay to zero and then rise to a value $f_0{\approx}0.7159$ around the time $T{=}6.282$, corresponding to the first revival. 

Another interesting feature of the revivals in Fig.~\ref{fig:2HC_dyn} is the presence of a small but visible peak at \emph{half} the revival period. In order to understand this, it is convenient to decompose the problem into a symmetric superposition of the two hypercubes and their antisymmetric superposition. The symmetric sector has $N{+}1$ states and its Hamiltonian is the one from Eq.~(\ref{eq:HCTB}) except that the last term of the sum is multiplied by a factor of $\sqrt{2}$. In the antisymmetric sector, the contribution of the two chains cancel at the middle vertex $\ket{000\ldots}$.  Therefore, this sector has only $N$ states and its Hamiltonian is the one of Eq.~(\ref{eq:HCTB}) without the last term in the sum. On their own, both sectors revive (although imperfectly) with a period close to $T{=}\pi$. However, at each revival the antisymmetric sector picks up a phase of -1. This is why the first revival of the full system only happens at $T{\approx} 2\pi$. The symmetric sector also has a slightly longer revival period that the antisymmetric one. The difference of frequency and amplitude of revivals means that they do not exactly cancel at $T{\approx} \pi$, hence the reflection peak.

Alternatively, this reflection can also be understood as being caused by reflection on the shared vertex "bottleneck".
See Appendix \ref{sec:2-HC} for more details.

\subsection{Interpolating between two hypercubes and the PXP model}

As shown in the previous section, the PXP model contains two embedded hypercubes of dimension $N/2$ which, on their own, support a revival from the $\ket{\mathbb{Z}_2}$ state in the thermodynamic limit. Here we explore a possible connection between the two-hypercube model and the full PXP model. Interpolation between the two models can be done naturally by varying the range of the projectors dressing the Pauli $X$ operator in Eq.~(\ref{eq:2hc}).  Specifically, the class of models interpolating between the PXP and two hypercubes are defined by 
\begin{equation}\label{eq:2cubeinterp}
H_{r} =\sum_jP_{j-2r+1}\cdots P_{j-3}P_{j-1}X_jP_{j+1}P_{j+3}\cdots P_{j+2r-1},
\end{equation}
where $r$ labels the number of projectors to the one side of $X$. Setting $r{=}1$ simply gives back the PXP model, whereas $r\geq N/4$ corresponds to the two-hypercube model in Eq.~(\ref{eq:2hc}).
As $r{=}0$ corresponds to the free spin-1/2 model, we can also consider this procedure as an interpolation between the two-hypercubes and the free spin-1/2 model where  PXP is just one of the intermediate steps.

\begin{figure}[htb]
\centering
\includegraphics[width=\linewidth]{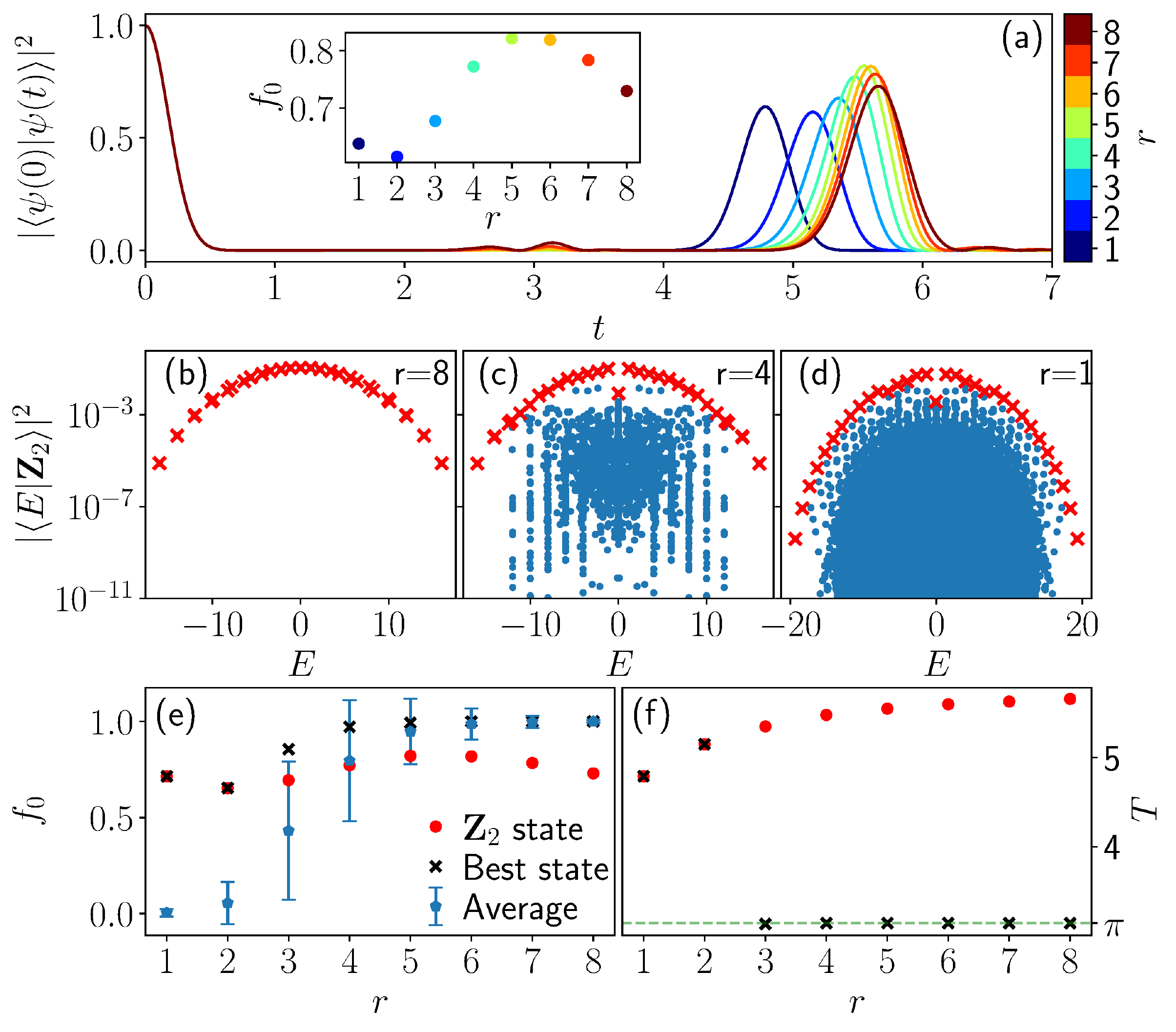}
\caption{ Revivals and scarred eigenstates in models defined in Eq.~(\ref{eq:2cubeinterp}) for different values of $r$. (a) Fidelity for the $\ket{\mathbb{Z}_{2}}$ initial state for $N{=}32$. The inset shows the fidelity at the first revival peak $f_0$ as a function of $r$. (b)-(d) Overlap between  $\ket{\mathbb{Z}_{2}}$  and the energy eigenstates for three different values of $r$ with $N{=}32$. Panel (b) corresponds to the two-hypercube model while (d) is PXP. The red crosses highlight the top band of $N{+}1$ eigenstates with anomalously high overlap with the $\ket{\mathbb{Z}_{2}}$ state. (e)-(f) $f_0$ and revival period $T$ for different $r$ values. In panels (e)-(f), we compare the revival period and the first revival peak $f_0$ for  $\ket{\mathbb{Z}_{2}}$ initial state with the computational basis state having the highest $f_0$ (``best state") as well as the average over all initial basis states (with the standard deviation shown by an error bar). The value of $N$ changes with $r$ in order to keep the Hilbert space dimensions comparable and in the range  $10^5{<}\mathcal{D}{<}1.5 \times 10^5$. 
}
\label{fig:IPX}
\end{figure}

All the models in Eq.~(\ref{eq:2cubeinterp}) have the two-hypercube as a subgraph and we compare the revivals from the N\'eel state in all of them in  Fig.~\ref{fig:IPX}(a) for a fixed value of $N{=}32$. We observe the fidelity at the first revival peak remains in the ballpark of $f_0{\sim} 0.7{-}0.8$ for all values of $r$, with a slight increase of the revival period with $r$. 
Moreover, for all values of $r$, we can identify a band of $N{+}1$ eigenstates with anomalously high-overlap on $\ket{\mathbb{Z}_{2}}$, see Fig.~\ref{fig:IPX}(b)-(d). The energy separation between these eigenstates is approximately constant in the middle of the spectrum and matches the frequency of revivals in Fig.~\ref{fig:IPX}(a).  For $r{=}N/4$, the Hamiltonian exactly corresponds to the two-hypercubes model and the spectrum contains only $N{+}1$ states. 
As $r$ is decreased,  this band of states evolves smoothly, while an increasing number of thermal eigenstates start to appear in the system -- see Fig.\ref{fig:IPX} (c) and (d).

In Fig.\ref{fig:IPX} (e) and (f) we compare the revivals from all states in the computational basis, i.e., we probe the revivals from \emph{all} graph vertices. To make a fair comparison between different models, instead of fixing the system size, we take a different value of $N$ for each model that gives roughly the same Hilbert space dimension. While for the N\'eel states there are few changes with $r$, this is not the case for most other initial states, whose revivals get worse as the constraint is relaxed. This can be understood by considering new vertices and edges that appear in the graph as $r$ is decreased.
In the rest of this work we will refer to these graph elements (vertices and corresponding edges) as ``bridges" as they are effectively bridging between the hypercubes.
When only the two hypercubes are present, the dynamics for the majority of states consists of state transfer in a single hypercube with a small leakage to the other hypercube. However, if a bridge is added close to a vertex, this will drastically enhance the state transfer to the other hypercube. Thus the dynamics is no longer well described by perfect state transfer with some leakage, and the revivals consequently get worse. On the other hand, for the N\'eel state the dynamics is relatively unchanged as we always have state transfer to the translated N\'eel state and back. Due to the form of the constraint, no bridge is added closer than at Hamming distance equal to 2 measured from that state.  Because of this, during the interpolation the dynamics is left relatively unchanged by the bridges. It also means that the first two steps of the FSA are identical and exact for all values of $r\geq 1$.

For all values of $r$, the top band of $N{+}1$ states is present and states belonging to it are decoupled from the bulk of the spectrum. These states can be well approximated by the FSA. As we change $r$, we see that the magnitude and period of the revival smoothly varies. These results suggest there is a form of ``adiabatic continuity" that protects the scarred subspace in the family of models in Eq.~(\ref{eq:2cubeinterp}). However, unlike the usual notion of adiabatic continuity, where the energy gap protects the smooth evolution of the ground state,  here we are looking at a subspace spanning a finite range of energy densities, which remains protected due to a combined effect of constraint and many-body scarring. 

\subsection{Alternative ways of enhancing the constraint}\label{sec:alt}

The persistence of revivals and scarred states as we vary $r$ in Eq.~(\ref{eq:2cubeinterp}) should be contrasted with perhaps a more intuitive way of strengthening the PXP constraint by simply increasing the Rydberg blockade radius:
\begin{eqnarray}\label{eq:ppxpp}
H_d = \sum_j \underbrace{PP\ldots P}_{d} X_j \underbrace{PP\ldots P}_{d}.
\end{eqnarray}
This family of models prevents an excitation in any contiguous block of $d{+}1$ sites and it has been realised in Rydberg atom arrays~\cite{Labuhn2016}. While $d{=}1$ reduces to the PXP constraint, for $d{>}1$ this clearly results in a different class of models compared to  Eq.~(\ref{eq:2cubeinterp}) above. We note that, similar to the PXP model, models with $d{>}1$ are also non-integrable and host a few exact many-body scarred eigenstates~\cite{SuraceConstrained}. The latter can be expressed in matrix product state form, analogous to PXP exact scars in Ref.~\onlinecite{Lin2019}. However, these exact scar states are not directly related to the band of $N{+}1$ scarred eigenstates illustrated in Fig.~\ref{fig:IPX} and they are unimportant for the revival dynamics that we focus on below. 

\begin{figure}[tb]
\centering
\includegraphics[width=\linewidth]{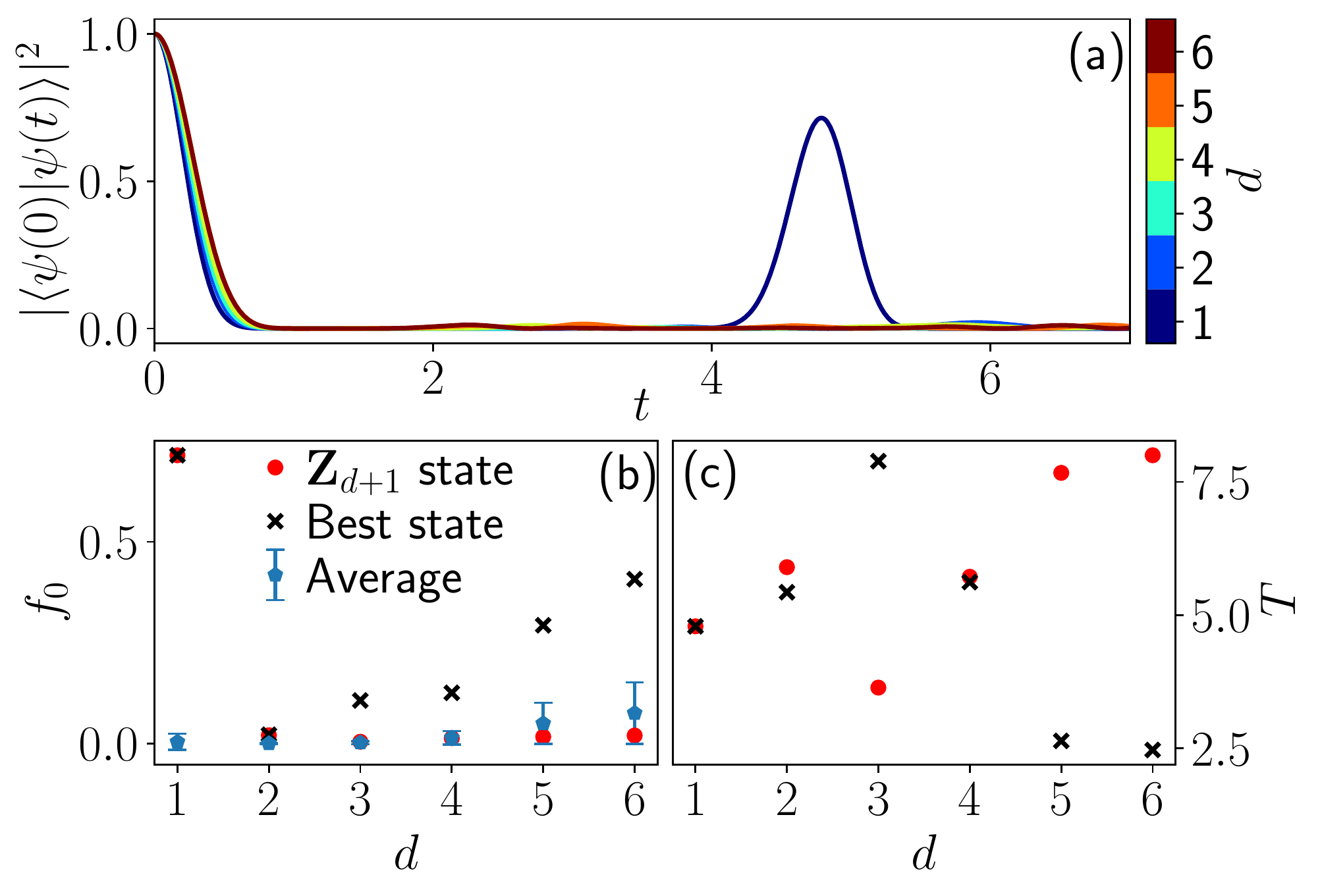}
\caption{ Revivals in models with a blockade radius $d$ described by Eq.~(\ref{eq:ppxpp}).  (a) The initial state   $\ket{\psi(0)} = \ket{\mathbb{Z}_{d{+}1}}$  shows no revivals for $d{>}1$. In panels (b)-(c), we compare the revival period $T$ and the first revival peak $f_0$ for the  $\ket{\mathbb{Z}_{d{+}1}}$ initial state with the computational basis state having the highest $f_0$ (``best state") as well as the average over all initial basis states (with the standard deviation shown as an error bar). In order to compare different models at approximately the same Hilbert space dimension, the system sizes are chosen according to the value of $d$ in such a way that the Hilbert space dimension is in the range $3.5\times 10^4{<}\mathcal{D}{<}1.1\times 10^5$. Each system size is also a multiple of $d{+}1$ in order for $\ket{\mathbb{Z}_{d{+}1}}$ to exist in the Hilbert space.
}
\label{fig:PXP_LR}
\end{figure}

In Fig.~\ref{fig:PXP_LR} we study dynamics for $H_d$ models in Eq.~(\ref{eq:ppxpp}). Because of the stronger constraint, models with $d{>}1$ do not contain the $\ket{\mathbb{Z}_2}$ state in their Hilbert space, hence we consider its generalisation 
\begin{eqnarray}\label{eq:zd}
\ket{\mathbb{Z}_{d+1}} = |1 \underbrace{0\ldots 0}_{d}1\underbrace{0\ldots 0}_{d}\ldots \rangle.
\end{eqnarray}
Fig.~\ref{fig:PXP_LR}(a) shows that only the PXP model with $d{=}1$ supports a discernible revival from the $\ket{\mathbb{Z}_{d{+}1}}$ states. While other initial states start to develop non-zero revival amplitude as $d$ is increased [Fig.~\ref{fig:PXP_LR}(b)], the revivals are generally worse than in the PXP model and the family of models studied in Fig.~\ref{fig:IPX}. 

To understand the lack of continuity between the models in Eq.~(\ref{eq:ppxpp}) as $d$ is varied, we need to identify the relevant subgraph associated with the initial states $\ket{\mathbb{Z}_{d{+}1}}$. Similar to the N\'eel state in all models described by Eq.~(\ref{eq:2cubeinterp}), $\ket{\mathbb{Z}_{d+1}}$ state holds the largest density of excitations in the corresponding model $H_d$ in Eq.~(\ref{eq:ppxpp}).
These excitations can be removed or added in any order as long as they are far enough from each other so that they are not affected by the constraint.
Consequently, each of the $d{+}1$ states obtained by translations of $\ket{\mathbb{Z}_{d{+}1}}$ is situated at one of the corners of a hypercube of dimension $N/(d{+}1)$, and these hypercubes form a ``star" pattern as they share a single vertex corresponding to the polarised state $\ket{000\ldots 00}$.
On their own, these $d{+}1$ hypercubes in a star configuration have good revivals from the $\ket{\mathbb{Z}_{d+1}}$ state, with state transfer to all the translated copies of $\ket{\mathbb{Z}_{d+1}}$ at half the revival period (see Appendix \ref{sec:n-cubes}).

However, the full Hamiltonian $H_d$ in Eq.~(\ref{eq:ppxpp}), has additional bridges in its graph that introduce connections between the hypercubes.
The effect of these bridges for different values of $d$ is hard to quantify, however  a relatively simple argument allows us to set the case $d{=}1$ apart from all $d{>}1$.
Indeed, just as PXP can be viewed as an intermediate step in the interpolation between the two-hypercubes and the free spin-1/2 model, the models with a Rydberg blockade of radius $d$ can be viewed as a step in the interpolation between $d{+}1$ hypercubes and the free spin-1/2 model. In order to understand the effect of the interpolation it is instructive to look a the dynamics at the two end points.
For $d{+}1$ hypercubes in a star pattern, the revivals occur because of state transfer between the state $\ket{\mathbb{Z}_{d+1}}$ and a symmetric superposition of all its translations.
Meanwhile, for the full hypercube, state transfer occurs between $\ket{\mathbb{Z}_{d+1}}$ and its opposite corner where all the spins are flipped: 
\begin{eqnarray}\label{eq:startransfer}
1 \underbrace{0\ldots 0}_{d}1\underbrace{0\ldots 0}_{d}\ldots \leftrightarrow  0\underbrace{1\ldots 1}_{d}0 \underbrace{1\ldots 1}_{d}\ldots
\end{eqnarray}
For $d{=}1$ this process is identical to swapping between the two N\'eel states related by translation.  For $d{>}1$, however, this is no longer true and  the dynamics must be significantly altered by bridges. The FSA  does not yield a good approximation for the dynamics away from the end points of the interpolation. For example, the FSA subspace dimension for $d{+}1$ hypercubes is $1{+}\frac{2N}{d+1}$ while for the free spin-1/2 model it is simply $N{+}1$.
It is clear from this alone that the subspace structure must change if $d{\neq }1$.

While this change in the dynamics could be better captured by an approximation scheme more complex than the FSA, numerical simulations of models with blockade radius $d$ do not show significant revivals for $d>1$ (see Fig.~\ref{fig:PXP_LR}). This shows that the presence of bridges in these models strongly affects the dynamics. This can be contrasted with the much weaker effect of bridges on the models studied in Fig.~\ref{fig:IPX}. In the latter case, the bridges added at each step do not change the dimension of the reviving FSA subspace. While these bridges still affect the revivals, they do it in a much less drastic way when their density is low, allowing a smooth transition from the two hypercubes to the PXP model. This suggests that the FSA is indeed the relevant approximation to capture the scarred dynamics, and that the failure of this approximation correlates very well with the absence of revivals.

\section{Weakening the constraint: interpolation between PXP and free spin-1/2 model via the $(k,k{+}1)$ models}\label{sec:kkp}

Instead of making the constraint stronger, one may wonder if by \emph{weakening} the constraint it might be possible to relate the many-body scarring in the PXP model with the free spin-1/2 model. This can be achieved by introducing a class of $(k,k{+}1)$ models with the constraint that each cell of $k{+}1$ sites can contain \emph{at most} $k$ excitations. The Hamiltonian for this series of models is given by the Rabi flip term compatible with the constraint, i.e., 
\begin{eqnarray}\label{eq:kmodels}
H_{(k,k+1)}= \mathcal{P}_k \left( \sum_j X_j \right) \mathcal{P}_k,
\end{eqnarray}
where $\mathcal{P}_k$ projects out any configuration with more than $k$ contiguous excitations anywhere in the chain.  Varying $k$ then allows to tune the effective strength of the constraint, with $k{=}1$ corresponding to the PXP model and $k{=}N$ being the free spin-1/2 model.

\begin{figure}[htb]
\centering
  \includegraphics[width=\linewidth]{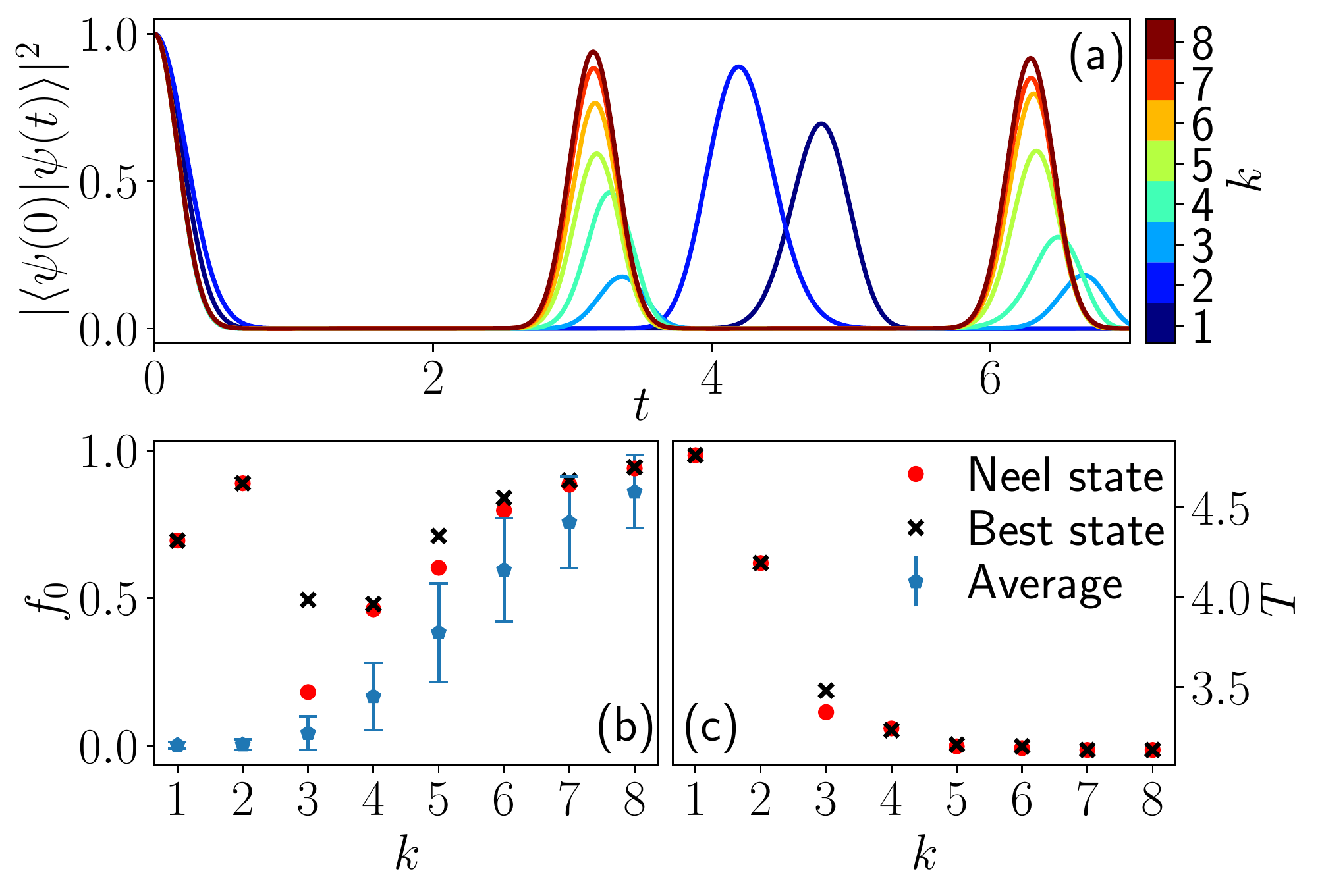}
\caption{ Revival fidelity for $(k,k{+}1)$ models  in Eq.~(\ref{eq:kmodels}). 
 (a) Fidelity time series for the initial N\'eel state and  various $k$.  (b)-(c) Fidelity at the first revival $f_0$ and the associated revival period $T$, for the initial N\'eel state, the best reviving basis state and the average over all computational basis states (with the standard deviation shown as an error bar). The discontinuity between $k{=}2$ and $k{=}3$ is clearly visible in the revival dynamics.  In order to compare different models at approximately the same Hilbert space dimension, the system sizes are chosen according to the value of $k$ in such a way that the Hilbert space dimension is in the range $1.3 \times 10^5<\mathcal{D}<2.8 \times 10^5$.
}
\label{fig:k_mods}
\end{figure}

Fig.~\ref{fig:k_mods} summarises the dependence of revivals in the models defined by Eq.~(\ref{eq:kmodels}) as a function of $k$. In the limit of large $k$, the behaviour is dominated by the proximity to the free spin-1/2 model, where many basis states revive. Intriguingly, we observe that the PXP model ($k{=}1$) is not smoothly connected to this large-$k$ limit. For example, the fidelity at the first revival peak, Fig.~\ref{fig:k_mods}(b), first increases in going from $k{=}1$ to $k{=}2$, but then drops precipitously from $k{=}2$ to $k{=}3$. The drop is sharp for the N\'eel initial state, but somewhat less pronounced if we look at \emph{all} initial basis states and choose the ``best" one. Nevertheless, this implies that scarred $\ket{\mathbb{Z}_2}$ dynamics in the PXP model cannot be understood by smoothly turning off the constraint to reach the free spin-1/2 model.  

From the FSA point of view, we expect $1{<}k\ll N$ models to support poorer revivals compared to the PXP model. Indeed, as more and more configurations are allowed, the graph starts to differ from the one of the two hypercubes as we get closer to the N\'eel state. This means that the FSA steps will start to become  inexact due to backscattering after fewer steps (see Appendix \ref{sec:fsa} for more details). For $k{>}3$, new states appear in the graph already in the first Hamming layer, and this is expected to strongly destabilise the revivals~\cite{Choi2018}. Similarly, for $k{=}2$ new states will appear in the second Hamming layer, in theory causing similar effects. However, this expectation is clearly not in agreement with Fig.~\ref{fig:k_mods} which shows that the $k{=}2$ model has more robust revival compared to the PXP model, for the same $\ket{\mathbb{Z}_2}$ initial state. In the remainder of this section, we study in detail the $k{=}2$ model and show that its scarring behaviour is a special case as it emerges from an underlying \emph{hypergrid} subgraph. This will serve as further evidence to the lack of continuity between the PXP model and the free spin-1/2 model, at least in the sense of  Eq.~(\ref{eq:kmodels}).

\subsection{Quantum many-body scars in the (2,3) model}\label{sec:k23}

The (2,3) model --  a special case of Eq.~(\ref{eq:kmodels}) where each consecutive triplet of sites can have at most two excitations -- bears many similarities with the PXP model. 
For example, we will show that the (2,3) model is non-integrable yet it hosts a band of $N{+}1$ scarred eigenstates with large support on the N\'eel state, $\ket{1010\ldots 10}$, reminiscent of the PXP model.  However, despite this similarity between the two models, we find the revivals and scarred eigenstates are more robust in the (2,3) model, even though the Hilbert space is larger in the latter model for the same size $N$. A more striking difference, which arises in sizes $N$ divisible by 4, is the existence of  additional reviving states, $\ket{11001100\ldots}$ and its three translated equivalents in the (2,3) model.

\begin{figure}[htb]
\centering
\includegraphics[width=\linewidth]{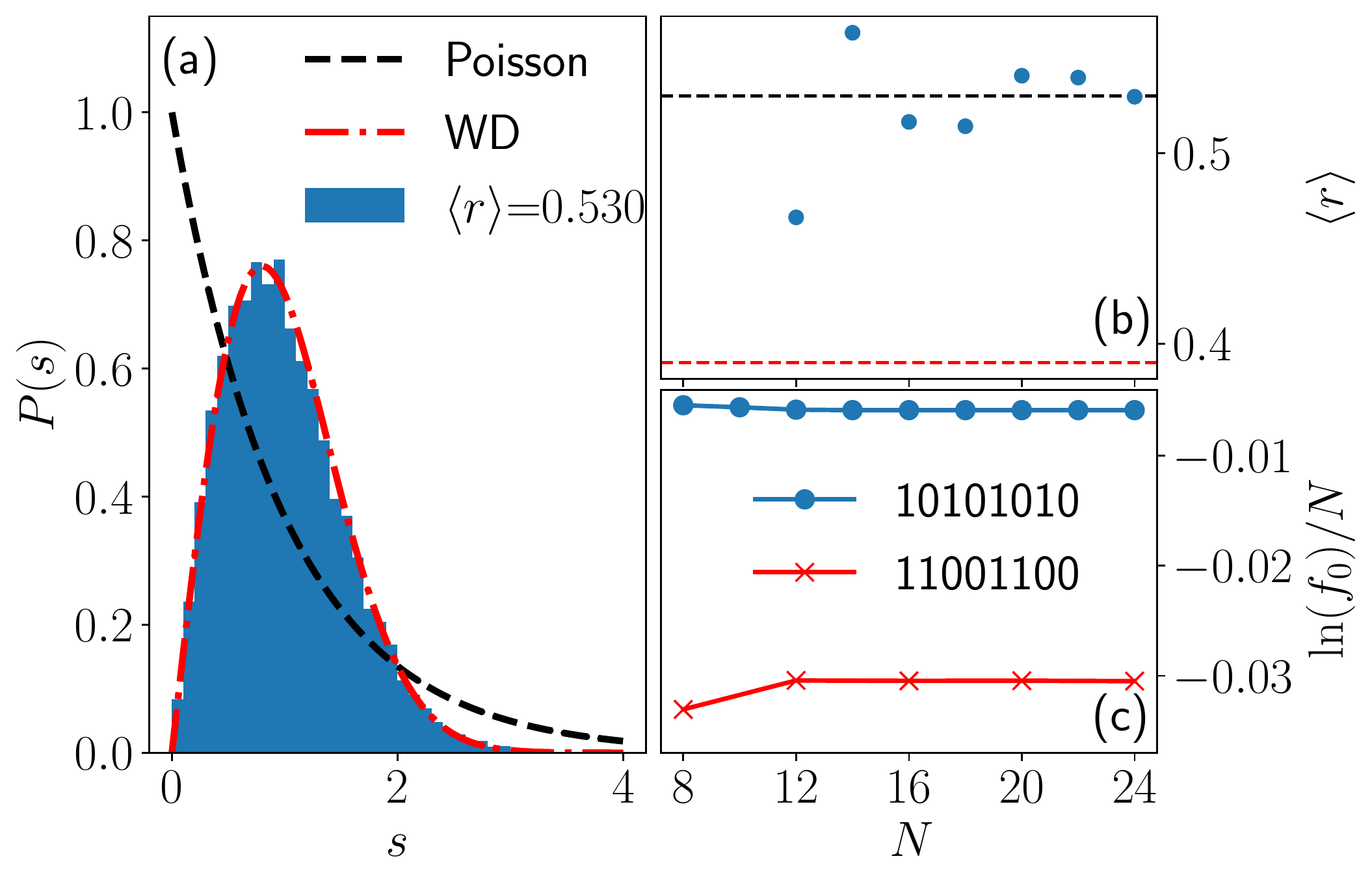}
\caption{
(a) Statistics of energy level spacings $P(s)$ in the (2,3) model in a large system size $N{=}24$ is consistent with the Wigner-Dyson ensemble. The average $r$-value is close to 0.53. The level statistics is computed in the momentum sector $K{=}0$ and inversion sector $I{=}{+}1$, after performing the spectrum unfolding. 
(b) Convergence of $\langle r \rangle$ with system size. (c) Fidelity density at the first revival from the N\'eel state in the (2,3) model saturates to a non-ergodic value -0.00591. For comparison, we also show the fidelity density of $|11001100\ldots\rangle$ initial state. Both of these fidelity densities are much larger than expected for a random initial state, signalling they are strongly atypical initial conditions for the (2,3) model.
}
\label{fig:23_level_stat}
\end{figure}

The (2,3) model is constrained and we first derive its quantum dimension, which determines the asymptotic scaling of the size of its Hilbert space. Following the method described in Appendix~\ref{sec:rec}, the Hilbert space dimension $\mathcal{D}_{N,k{=}2}$ for the (2,3) model obeys the recurrence relation 
\begin{equation}
\mathcal{D}_{N,2}=\mathcal{D}_{N-1,2}+\mathcal{D}_{N-2,2}+\mathcal{D}_{N-3,2},
\end{equation}
for $N{>}3$, with $\mathcal{D}_{1,2}{=}1$, $\mathcal{D}_{2,2}{=}3$ and $\mathcal{D}_{3,2}{=}7$ for PBC. 
The quantum dimension $\alpha_2$ must satisfy $\alpha_2^3-\alpha_2^2-\alpha_2-1=0$, which gives 
\begin{equation}
\alpha_2{=}\frac{1}{3}\left(1 {+} \sqrt[3]{19 {-} 3 \sqrt{33}} {+} \sqrt[3]{19 {+} 3 \sqrt{33}}\right){\approx}1.839.
\end{equation}
Thus, the dimension of the Hilbert space of the (2,3) model grows asymptotically as ${\sim}1.8^N$.

We next demonstrate that the (2,3) model is non-integrable and it supports revivals due to the existence of $N{+}1$ towers of scarred eigenstates. The average energy level spacing $\langle r \rangle $~\cite{OganesyanHuse}  is found to approach 0.53 in large systems, see Figs.~\ref{fig:23_level_stat} (a)-(b), as expected from a thermalising system. Moreover, the distribution of energy level spacings is consistent with the Wigner-Dyson ensemble~\cite{Mehta2004}, demonstrating that physical properties of the model cannot be explained by its proximity to the full hypercube.

The fidelity density at the first revival, $\ln(f_0)/N$, is computed  for the N\'eel and $|11001100\ldots\rangle$ initial states and several values of $N$ in Fig.~\ref{fig:23_level_stat}(c).  For the N\'eel state, the fidelity density quickly saturates to ${\approx}{-}0.00591$, indicating weak finite-size effect. For a random initial state one would expect the saturation value to be $-\text{ln}(\alpha_2) {\approx}{-}0.609$.
As this is two orders of magnitude larger than the actual value,  it shows that the revivals are not simply fluctuations due to a small finite size of the system. For comparison, we also study the other reviving state,  $\ket{1100\ldots 1100}$, whose fidelity density converges to ${\approx}{-}0.0304$. While larger than for the N\'eel, this value is still an order of magnitude smaller than for a random state, signalling that this initial state is also atypical for the (2,3) model. 

\begin{figure}[htb]
\centering
    \includegraphics[width=0.92\linewidth]{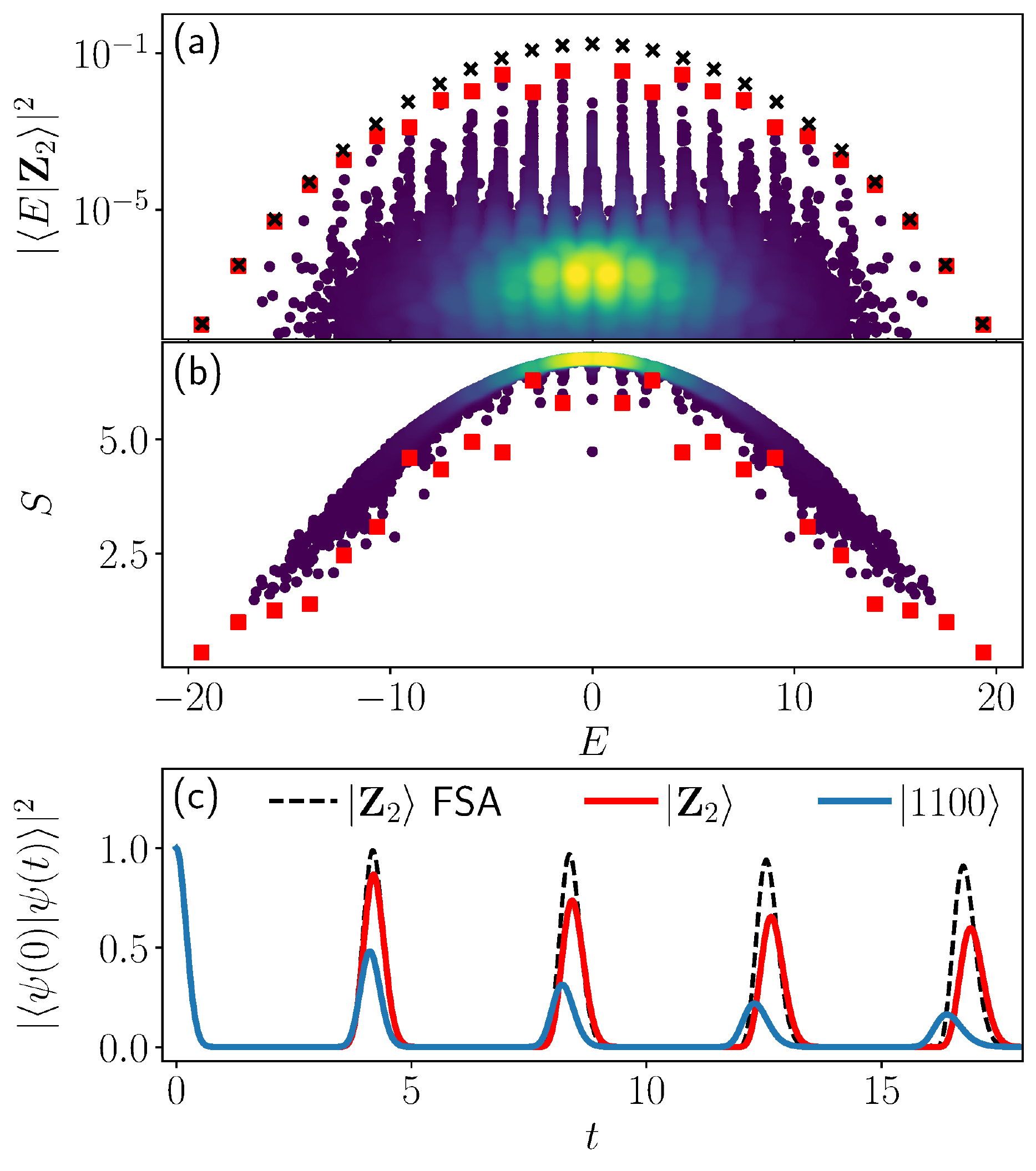}
\caption{Dynamics and eigenstate properties in the (2,3) model. 
(a) Overlap of eigenstates with $|\mathbb{Z}_2\rangle$ state.
 $N{+}1$ scarred eigenstates with anomalously high overlap are labelled by red squares. The FSA approximates well the energies of  scarred eigenstates (black crosses).
 (b) Entanglement entropy of all eigenstates in momentum sectors $K{=}0$ and $K{=}\pi$, with the same $N{+}1$ scarred eigenstates highlighted in red. 
 (c) The FSA provides a good estimate of the revival frequency from $|\mathbb{Z}_2\rangle$ initial state, although it somewhat overestimates the revival amplitude. For comparison, we also show the dynamics from  $\ket{1100 \ldots}$ initial state. 
 All data is for system size $N{=}24$. In (a) and (b) the color scale indicates the density of data points.
}
\label{fig:23_olap_S}
\end{figure}

In Fig.~\ref{fig:23_olap_S}(a) we study the overlap between the reviving  N\'eel state and the eigenstates of the (2,3) model. Similar to the PXP model studied in Ref.~\onlinecite{TurnerPRB}, we observe that eigenstates form tower structures. At the top of each tower is a scarred state with high overlap on the N\'eel state. The FSA subspace provides a very good estimate of the energy of each tower, as indicated by crosses in Fig.~\ref{fig:23_olap_S}(a). The FSA also captures the revival dynamics, as shown in Fig.~\ref{fig:23_olap_S}(c), in particular it accurately estimates its frequency, while somewhat overestimating the amplitude of the revival. 

Moreover, scarred eigenstates can also be identified as having much lower entanglement than other eigenstates at the similar energy density. To quantify entanglement, we compute the von Neumann entanglement entropy, $S{=}{-}\mathrm{Tr} \rho_{N/2} \ln \rho_{N/2}$, where $\rho_{N/2}$ is the reduced density matrix for one half of the chain. Entanglement entropy of eigenstates of the (2,3) model in momentum sectors $K{=}0$ and $K{=}\pi$ is shown in Fig.~\ref{fig:23_olap_S}(b). Entanglement entropy reveals the  scarred eigenstates as some of the most weakly entangled states in the spectrum. While in smaller systems entropy distribution shows a large spreading, similar to the PXP model as pointed out in Ref.~\onlinecite{Khemani2018}, in larger system sizes like $N{=}24$ shown in Fig.~\ref{fig:23_olap_S}(b), we observe that entropy distribution becomes quite narrow, starting to look more similar to models such as AKLT~\cite{BernevigEnt} and constrained clock models~\cite{Bull2019}. In particular, constrained clock models exhibit very narrow towers densely populated with eigenstates, as also seen in Fig.~\ref{fig:23_olap_S}(a). Such towers enhance the hybridisation between the top $N{+}1$ scarred eigenstates and the rest of the spectrum, resulting in relatively high entropy of scarred eigenstates,   Fig.~\ref{fig:23_olap_S}(b). 

\subsection{Hypergrid subgraphs in the (2,3) model}\label{sec:23hcl}

The key difference between the PXP and (2,3) models can be traced to the underlying subgraph associated with the revivals from the N\'eel and $|11001100\ldots\rangle$ initial states. While in the PXP and other models studied up to this point the relevant subgraph was a union of hypercubes sharing a single vertex (Sec.~\ref{sec:pxp_2cube}), in the (2,3) model we find a different type of subgraph consisting of two \emph{hypergrids} of dimension $N/2$. In this section, we provide analytical and numerical evidence that the hypergrid subgraphs are indeed responsible for the atypical dynamics and many-body scarring in the (2,3) model.

\subsubsection{Proof of the existence of two hypergrids in the (2,3) graph}

A hypergrid graph $\HG{m}{d}$ is defined as the Cartesian product 
\begin{equation}
\HG{m}{d}
=\underbrace{L_{m} \square L_{m} \square \ldots \square  L_{m}}_d,
\end{equation}  
where $L_{m}$ stands for a linear graph of order $m$ (with $m$ vertices) and we are only interested in hypergrids with the same order in all dimensions. 
For example, the hypergrid $\HG{2}{d}$ is simply the hypercube of dimension $d$. Similarly, the hypergrid graph $\HG{2S+1}{d}$, having $2S+1$ states in each dimension, is isomorphic to an unweighted graph of a free spin model with $d$ spin-$S$ degrees of freedom. This is because each vertex of $\HG{m}{d}$ can be labelled by a $m$-ary string $\{1,2, \ldots m \}^d $, and only vertices with a single site differing by 1 are connected by an edge.  Note that for $S{>}1$ the matrix elements of the free spin-S model Hamiltonian are no longer equal, and the model can no longer be described solely by an unweighted graph. We will not consider such cases in this paper.

\begin{figure}[tb]
\centering
\includegraphics[width=0.5\linewidth]{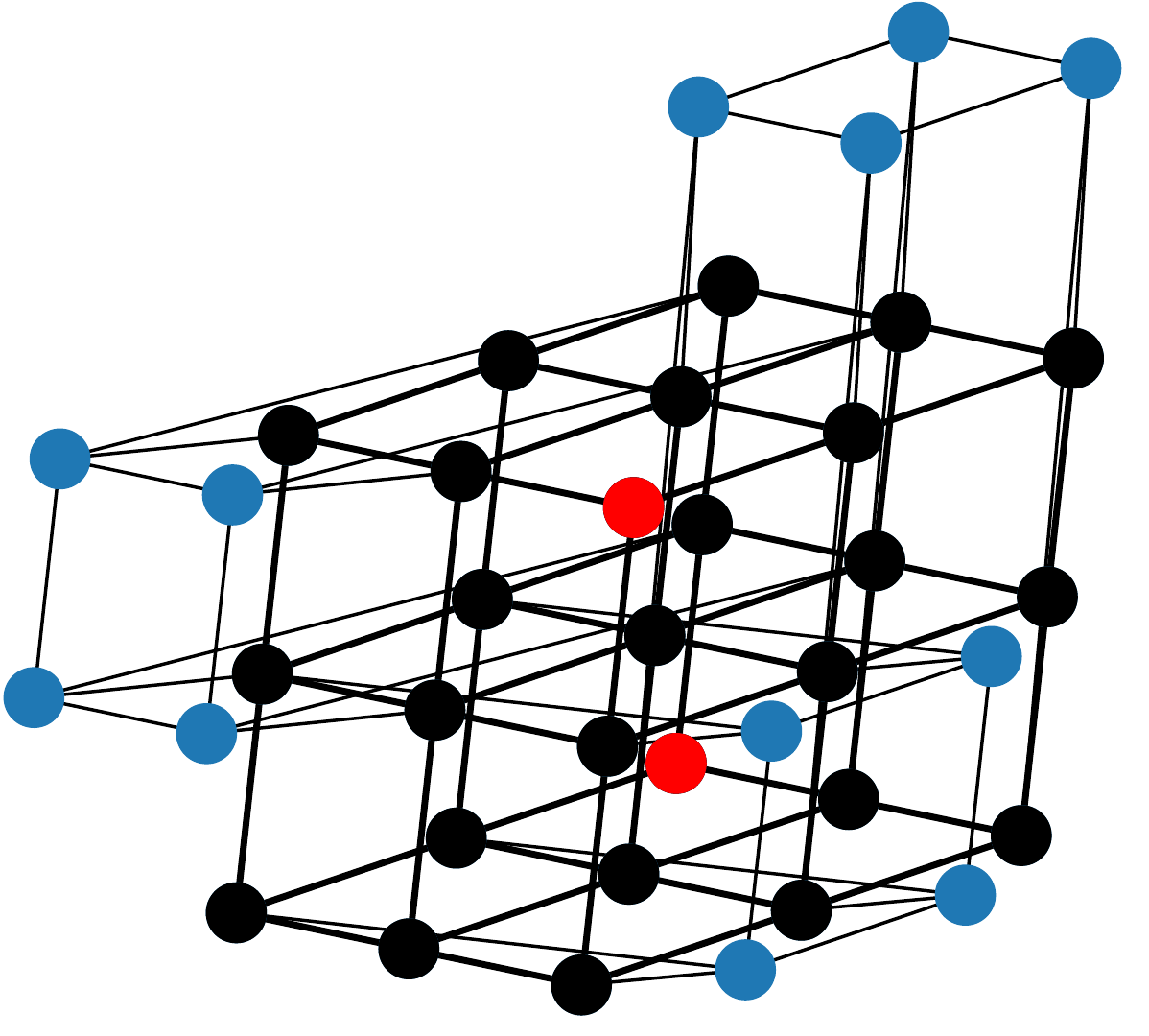}
\caption{Adjacency graph of the (2,3) model for system size $N{=}6$. The subgraph associated with many-body scarring is outlined in black, while the N\'eel states are shown in red. Blue vertices denote bridges connecting vertices in the subgraph. The subgraph is isomorphic to a hypergrid $\HG{3}{N/2}$. Note that there are two such hypergrids, but for clarity only one of them is highlighted here.
}
  \label{fig:23_HCL}
\end{figure}

Next we show that there are two distinct hypergrids $\HG{3}{N/2}$ that can be identified as subgraphs of the (2,3) model. 
One of these hypergrids is sketched in Fig.~\ref{fig:23_HCL}.
The proof is based on grouping sites into pairs~\cite{Lin2019,Shiraishi_2019}. Let us first define the states $\ket{o}=\ket{00}$, $\ket{L}=\ket{10}$, $\ket{R}=\ket{01}$, $\ket{2}=\ket{11}$.
In this formulation, the only forbidden configurations are $\ket{2L}$, $\ket{R2}$, and of course $\ket{22}$. The Hamiltonian acting on the $N/2$ pairs of sites can be written as
\begin{equation}
H_{(2,3)}=\sum_{b=1}^{N/2}h_{b-1,b,b+1},
\end{equation}
where the local Hamiltonian term is
\begin{equation}\label{eq:H_23_cell}
\begin{split}
h_{b-1,b,b+1}&=\id\otimes \big(\ket{o}\bra{R}+\ket{R}\bra{o}\big)\otimes \big( \id-\ket{2}\bra{2}\big) \\
&+\big( \id-\ket{2}\bra{2}\big) \otimes \big(\ket{o}\bra{L}+\ket{L}\bra{o}\big)\otimes \id \\
&+\big(\ket{o}\bra{o}{+}\ket{L}\bra{L}\big){\otimes} \Big[\ket{L}\bra{2}{+}\ket{2}\bra{L} \\
&+\ket{R}\bra{2}{+}\ket{2}\bra{R}\Big]{\otimes} \big(\ket{o}\bra{o}{+}\ket{R}\bra{R}\big).\\
\end{split}
\end{equation}

Let us take $N{=}8$ as an example. Start from the N\'eel state $\ket{\mathbb{Z}_2}=\ket{10101010}$ and group the cells into  pairs $(1,2)$, $(3,4)$, $(5,6)$ and $(7,8)$.
Then $\ket{\mathbb{Z}_2}=\ket{LLLL}$ and every pair of sites can be freely flipped $\ket{L}\rightleftarrows\ket{o}\rightleftarrows\ket{R}$, like a spin 1.
This means that there is a hypergrid graph $\HG{3}{4}$ between the N\'eel state $\ket{LLLL}$ and the anti-N\'eel $\ket{RRRR}$.
It is important to note that while in the $(2,3)$ model under some condition the flips $\ket{L}\rightleftarrows\ket{2}$ and $\ket{R}\rightleftarrows\ket{2}$ are possible, they correspond to bridging out of this hypergrid graph.
Beyond the N\'eel states, we would expect to also see revivals from other corners of this hypergrid, i.e., from states in which all cells have an extremal value (either $L$ or $R$). Indeed, this would mean that all cells would  precess freely with the same frequency.
However, the only other corners of this graph that have no edges going out of the hypergrid are $\ket{LRLR}=\ket{10011001}$ and $\ket{RLRL}=\ket{01100110}$. Indeed, from Eq.~(\ref{eq:H_23_cell}) one can see that any $LLR$ or $LRR$ configuration can be changed to an $L2R$ one which is not in the hypergrid.

Alternatively, the sites can be paired up as $(8,1)$, $(2,3)$, $(4,5)$ and $(6,7)$. 
In this case the N\'eel state is $\ket{RRRR}$ and the same spin-1 argument holds. 
However, this hypergrid graph $\HG{3}{4}$ is different from the last one, as can be seen by looking at the corners $\ket{LRLR}=\ket{00110011}$ and $\ket{RLRL}=\ket{11001100}$. 
In the first formulation these states would be $\ket{o2o2}$ resp. $\ket{2o2o}$, which are not in the corresponding hypergrid graph.

The two hypergrids identified above are not equivalent but share several vertices, and their union can be taken as a model on its own, which we refer to as ``2HG" model.
In fact all states with no neighbouring excitations belong to both hypergrids, so their intersection gives back the PXP graph.
Because of this,  the total number of states in the 2HG model is asymptotically given by $3^{N/2}-\phi^N$, where $\phi$ is the Golden Ratio.

A single hypergrid has perfect state transfer and revivals from any corner state. While the revivals in the 2HG graph are no longer perfect, they are still present with a similar frequency. 
The two N\'eel states are corners of both hypergrids and they are the best reviving basis states in the (2,3) model, while the states $\ket{110011\ldots 1100}$ are all corner of only one of the hypergrids and their revivals are found to be weaker, as expected from their position in the graph. 
All other basis states are either not corners of these hypergrids or they have additional edges extending  out of the 2HG, and thus they are not expected to revive.
Finally, if $N$ is even but not a multiple of 4, the two N\'eel states are the only reviving ones, as all other corners of the hypergrid have edges going outside of it.

\subsubsection{Numerical evidence for the relevance of hypergrids for many-body scarring}

\begin{figure}[bt]
\centering
\includegraphics[width=0.9\linewidth]{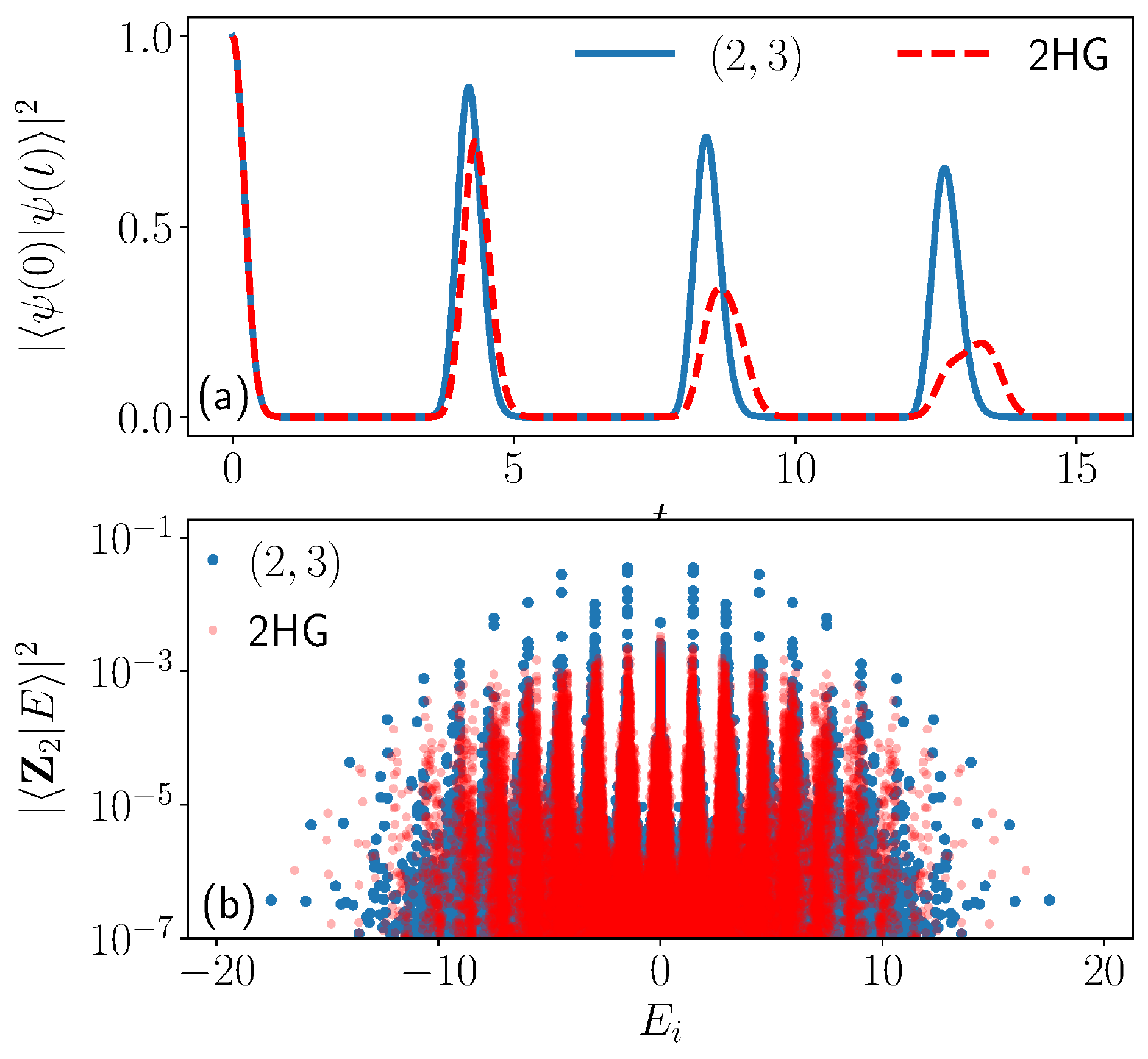}
\caption{Scarred dynamics and eigenstate properties in the (2,3) model and its 2HG subgraph. (a) The revivals from the N\'eel state have very close frequency in the two models, but their decay is more pronounced in 2HG model.  This can be attributed to the absence of a well-separated band of scarred eigenstates for this model, as can be seen in (b). Data is for system size  $N{=}24$.
}
\label{fig:23_comp}
\end{figure}

In Fig.~\ref{fig:23_comp}, we numerically test the relevance of the 2HG subgraph for  many-body scarring in the (2,3) model. We compare the dynamics and eigenstate properties in the (2,3) model with their projection into the 2HG model. 
In both models, we observe revivals of the wave function, with similar frequencies,  see Fig.~\ref{fig:23_comp}(a). However, the amplitude of revivals decays more rapidly in the 2HG model compared to the (2,3) model. This difference can be related to the eigenstate overlap with the N\'eel state shown in Fig.~\ref{fig:23_comp}(b). 
The overlap between the N\'eel state and the eigenstates of the (2,3) model shows clear tower structures with an energy spacing close to that in the 2HG model. However, in the 2HG model there is no top band of states that is well-separated from the bulk like in the (2,3) model or the PXP model in Fig.~\ref{fig:IPX}. Thus, the revivals decay faster as more states participate in the dynamics. 

The hypergrids also seem to play an important role in stabilising the first step of the FSA. Indeed, because of the constraint it is possible to add excitations to a state already in the second Hamming layer. This means that only the first step of the FSA is the same as in the two-hypercube model. Based on that we would expect to only have a single exact FSA step and thus poor revivals as in (3,4) and (4,5) models (see Appendix \ref{sec:fsa} for more details). However, in practice we observe that the first two FSA steps are exact, as is also the case in the 2HG model.

The bridges added on top of 2HG to form the (2,3) model seem to stabilise the revivals from the N\' eel state. The exact mechanism by which this happens is unclear to us, however we believe that the mechanism is non-generic as the addition of random bridges is found to consistently lead to poorer revivals. At the same time, the revivals from most other states are destroyed by additional bridges. As we demonstrate in the next Section, this mechanism of revival stabilisation due to a small density of bridges is also realised in the two-hypercube model describing the many-body scarring in the PXP model.

\section{Connecting two hypercubes via bridges}\label{sec:pxp_bridges}

In Secs.~\ref{sec:pxp_2cube} and \ref{sec:kkp} we studied two classes of models with the structure of partial cubes which were shown to contain hypergrid subgraphs. We now turn towards building new constrained models directly from their graph. We start from  two hypercubes and add back states from the unconstrained spin-1/2 chain of the same length.
All models sampled in this procedure will have the form of Eq.~(\ref{eq:p_cube}), with the additional constraint that $\mathcal{P}$ only prevents exciting new sites and never removing excitations.
Beyond this property, all models sampled will also be invariant under translation. Thus, when we grow the graph, a new basis state with $m$ excitations is randomly chosen. After that, all vertices corresponding to this state, its translations, or states that can be obtained by removing excitations from these, are added along with the relevant edges. 
After each addition, the first revival and the revival period are computed.
The value of $m$ is initialised at 2 and increased after a step if some conditions are met. Details of the algorithm can be found in Appendix~\ref{sec:algo}.

In order to monitor closeness to the two hypercubes or to the free spin-1/2 model, we introduce the bridge-density parameter $\lambda$ defined as
\begin{equation}\label{eq:bridgedensity}
\lambda=\frac{\text{ln}(|G|)-\text{ln}(2^{N/2+1}-1)}{\text{ln}(2^N)-\text{ln}(2^{N/2+1}-1)},
\end{equation}
 where $|G|$ is the number of states in the graph. 
For the two hypercubes joined at a vertex, $\lambda{=}0$, while for a single large hypercube $\lambda{=}1$.  The logarithms ensure that $\lambda$ is properly normalised in large systems. 

\begin{figure}[tb]
\centering
\includegraphics[width=\linewidth]{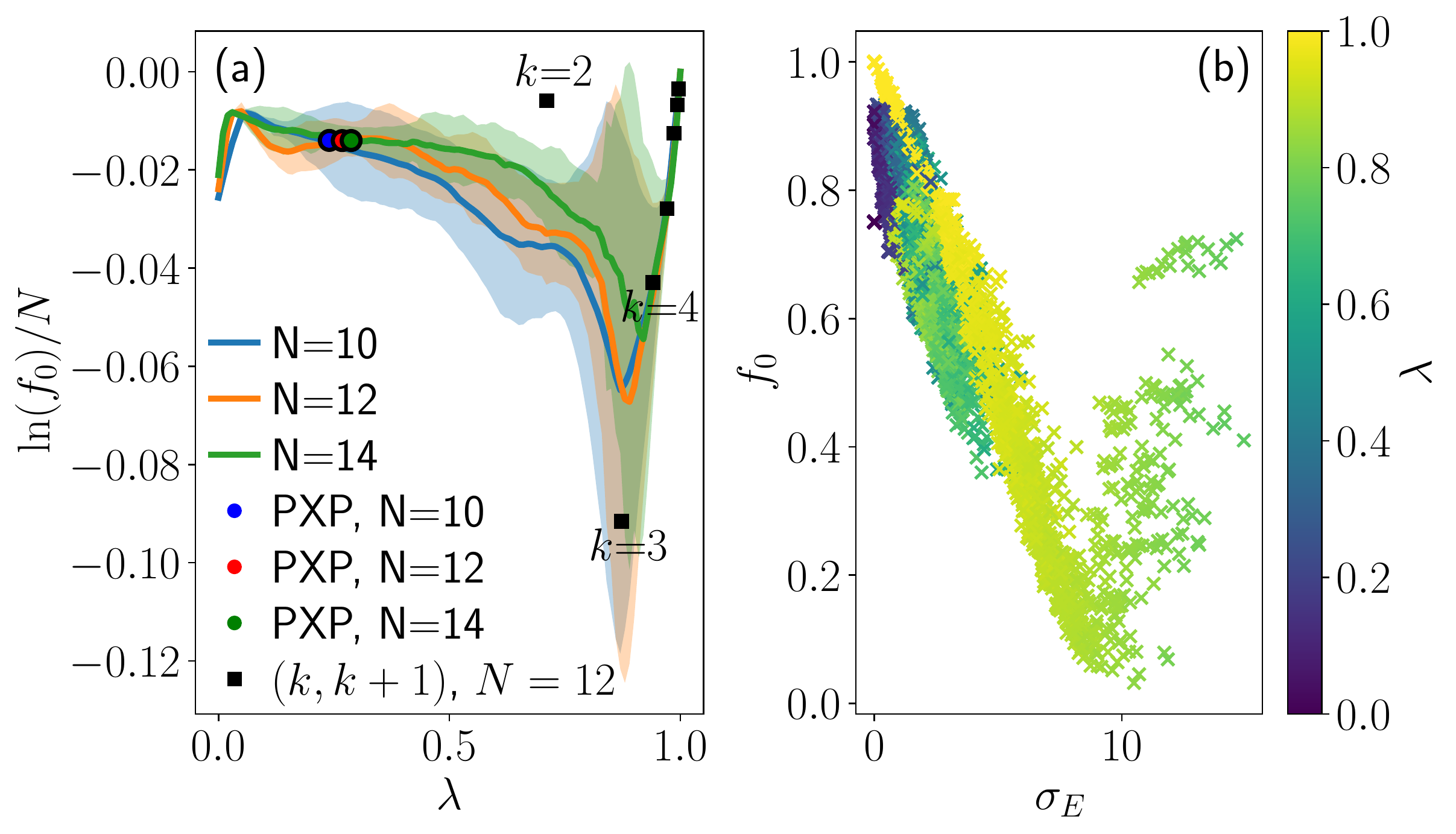}
\caption{(a) Revival fidelity density when random bridges of density $\lambda$ are added to the two-hypercube model in a few system sizes $N$.
The shaded areas represent standard deviation over different realisations of the bridges with the given density. This analysis reveals that the PXP model falls in the middle of the distribution, thus it is a ``generic" model with the given density of bridges. The behaviour of most $(k,k{+}1)$ models is also close to the expected average, with $k{=}2$ being a notable outlier.
(b) Revival fidelity for random bridges added to the two-hypercube model with $N{=}12$. The subspace variance of the FSA, $\sigma_E$, correlates well with the fidelity at the first revival. }
\label{fig:2cubes_bridge}
\end{figure}
The result of adding bridges is summarised in Fig.~\ref{fig:2cubes_bridge} for  chains of length $N{=}10, 12, 14$. 
Fig.~\ref{fig:2cubes_bridge} shows that the PXP model represents a typical model with the given density of bridges added to the two hypercubes. The presence of the two hypercubes explains why this model revives, and the bridges only weakly affect the fidelity and period of the revivals.
For larger values of $k$, the corresponding $(k,k{+}1)$ models generally fall very close to the average of random models for the same value of $\lambda$ due to their proximity to the free spin-1/2 model. A notable exception is $k{=}2$, which has significantly better revivals than expected from our random sampling analysis. As we argued previously, this is likely due to the presence of the 2HG subgraph and the special structure of the bridges in that model.

Another notable feature of Fig.~\ref{fig:2cubes_bridge} is that there is always an improvement of the revivals when a small number of bridges are added to the two hypercubes. Intuitively, we would expect the fidelity to decay as the graph gets further away from the two hypercubes, until it becomes  close enough to the full hypercube of a larger size, which also has good revivals. The enhancement of revivals at low values of $\lambda$ can be  understood as ``correcting" the frequency mismatch between the symmetric (resp. antisymmetric) superpositions of the two hypercubes, as we discussed in Sec.~\ref{sec:twocubes}.
In this regime the bridges only affect the frequency of the symmetric sector, bringing it closer to the frequency of the antisymmetric sector, thus improving the revival fidelity (see  Appendix \ref{sec:2-HC} for further details).
It is also important to note that the range of $\lambda$ where this improvement happens goes to zero in the thermodynamic limit, meaning that the slope of the curve in the limit $\lambda{\to} 0$ in Fig.~\ref{fig:2cubes_bridge}(a) becomes steeper with the increase in system size.  

For all graphs sampled during the process in Fig.~\ref{fig:2cubes_bridge} the dimension of the FSA subspace remains unchanged.
Indeed, the FSA process from the N\'eel states always terminates on the anti-N\'eel state after $N$ steps, leading to $N{+}1$ states.
In addition, for the two joined hypercubes and for the full hypercube this subspace is exact, meaning that it is disconnected from the rest of the Hilbert space. For the random graphs sampled, the FSA is generally not exact, and this can be quantified using the subspace variance $\sigma_E$, as explained in Appendix~\ref{sec:fsa}.
Among all graph properties, the subspace variance was found to best correlate with the existence of revivals, see Fig.~\ref{fig:2cubes_bridge} (b). This correlation implies that the FSA revivals are generally good, and that the leakage out of it is the main factor that destabilises the revivals in the full system.
As the algorithm adds back states with incrementally more excitations, these new vertices can get closer and closer to the N\'eel state.
This means that they can affect the FSA at earlier steps and thus exert a stronger effect on the revivals.

We also performed the sampling procedure of adding bridges to three and four hypercubes $\HG{2}{N/3}$ (resp. $\HG{2}{N/4}$), all joined at a single vertex.
These structures are found in the longer-range Rydberg blockades models in Eq.~(\ref{eq:ppxpp}). In contrast to two hypercubes, the models with $\lambda{\approx} 0.5$ showed very poor revivals.
We understand this difference as emanating from a much bigger change of the FSA subspace.
Indeed, for the two cubes the FSA subspace always has dimension $N{+}1$, whereas for more hypercubes this dimensions changes from $2N/3{+}1$ (resp. $N/2{+}1$) to $N{+}1$ as more bridges are added.
More details can be found in Appendix \ref{sec:n-cubes}.

\section{Conclusions and discussion}\label{sec:conc}

In this paper we explored a possible origin of many-body scars and associated  wave function revivals in the PXP model describing arrays of Rydberg atoms. We studied the properties of the Hamiltonian adjacency graph, in particular the existence of large regular subgraphs, as we varied the constraint in the PXP model. We considered two simple limiting cases, the free spin-1/2 model and the model of two hypercubes joined at a single vertex, which naturally arise when the constraint is either completely turned off or made stronger to penalise not only nearest-neighbour excitations, but an entire sublattice of the chain. While both of these limits support revivals in the thermodynamic limit, we argued that only the two hypercube model faithfully captures the many-body scarring phenomenology in the PXP model. To demonstrate the connection between the two, we introduced a family of models with a variable range of the constraint, showing that the scarred subspace remains preserved under this interpolation. By contrast, such a smooth interpolation was not found between PXP and the free spin-1/2 model. Nevertheless, the exploration of this connection led us to new constrained models, such as the (2,3) model, which were shown to have unique scarring phenomenology of their own. 

We note that the Hamiltonian adjacency graph has recently been linked to quantum many-body scars in a few models~\cite{Desaules2021,surace2021quantum,yoshinaga2021,bosonScars,Voorden2021}. However, these studies focus on regular subgraphs with  \emph{weak} connectivity to the the rest of the Hilbert space. In contrast, the subgraphs identified in this work do not have this property. For example, in the PXP and (2,3) models, the bridges form an essential part of the scarred dynamics and even enhance it, as opposed to simply destabilising revivals. 

Our analysis suggests that large families of scarred quantum networks can be generally built according to the following three steps. (i) We start from a highly-structured graph obtained by joining two or more fundamental components, each of which individually features perfect state transfer. One example of this is two hypercubes joined at a single vertex, but other arrangements are also possible, such as more than two hypercubes (e.g., a linear array) or a union of hypergrid graphs. The composite graph should retain robust, if imperfect, revivals. In several examples studied in this paper, the composite graph in fact has many reviving vertices -- for example, in the two-hypercube model, many reviving vertices exist away from  the axis which passes through  the two N\'eel corners and the joint vertex ($000\ldots$), recall  Fig.~\ref{fig:graph_struct}. (ii) Next, we add bridges that connect the fundamental components. These bridges will typically destroy the revivals from many of the vertices, but they can stabilise the revival in a small number of the remaining vertices. (iii) The final graph defines a quantum model. A typical caveat is that this model is not necessarily expressed in terms of a local Hamiltonian, but we showed that there are examples resulting in a local model. A notable example of this is the PXP model, where the bridges between the two hypercubes are realised by a simple local constraint on neighbouring excitations.
Our approach sheds light on the relation between constrained systems and many-body scarring. Indeed, constraints have the effect of removing vertices and edges from the graph. If the constraint removes enough bridges while leaving the substrucre intact, then it can create the right conditions for scarring and revivals.

In much of the existing literature, quantum many-body scars and other kinds of non-stationary dynamics have been understood from the su(2) algebra point of view, where the non-thermalising eigenstates form a representation of the algebra~\cite{BernevigEnt, Buca2019, Iadecola2019_2, MotrunichTowers, Buca2019_2, Dea2020}. This eigenstate-based picture, however, does not directly allow to predict  the existence of many-body scars without diagonalisation of the  Hamiltonian -- an exponentially difficult task. Our approach instead  focuses on the Hamiltonian matrix and its properties in the computational basis. While in general we expect there is no easy way to  directly relate the two points of view, we found that in many scarred models the existence of regular subgraphs, judiciously perturbed by bridges,  correlates with the emergence and enhancement of su(2) algebra, as captured by the forward scattering approximation. While our analysis has been primarily numerical, it would be interesting to analytically affirm  the connection between the emergent su(2) algebra and the underlying regular subgraph in future work.

{\sl Note added:} During the completion of this manuscript, we became aware of Ref.~\onlinecite{dooley2021extreme} which also investigated some of the constrained models introduced here, and of Ref.~\onlinecite{windt2021squeezing} which solved the two-cube model analytically in the thermodynamic limit. Our results are in agreement where they overlap. 

\subsection*{Acknowledgements}
We thank Christopher Turner and Maksym Serbyn for useful discussions. We acknowledge support by EPSRC grant EP/R513258/1 and by the Leverhulme Trust Research Leadership Award RL-2019-015.
Statement of compliance with EPSRC policy framework on research data: This publication is theoretical work that does not require supporting research data.

\appendix

\section{Forward scattering approximation for other models} \label{sec:fsa}

For the model of two hypercubes joined at a vertex, the PXP model and the $(k,k+1)$ model, the FSA scheme introduced in Sec.~\ref{sec:fsa1cube} needs to be modified  to include the constraint. Once again, we start the FSA from  $ \ket{v_0}=\ket{\mathbb{Z}_2}$ state but  we redefine the forward and backward propagating parts.
For all models that can be written as Eq.~(\ref{eq:p_cube}), it follows:
\begin{align}\label{Eq:Hpm}
H^{\pm} &= {\sum_{j \in \text{ even}}} \mathcal{P}\sigma^\pm_j\mathcal{P}  + {\sum_{j\in \text{ odd}}} \mathcal{P}\sigma^\mp_j \mathcal{P},
\end{align}
with $\mathcal{P}$ the global constraint for the particular model studied.
Similar to the case of the free spin-1/2 model in Sec.~\ref{sec:fsa1cube}, in such a decomposition $H^+$ always increases the Hamming distance from the N\'eel state and $H^-$ always decreases it, and their sum corresponds to the full Hamiltonian. In the graphs of the two-hypercube and the PXP models, $|\mathbb{Z}_2\rangle$ state is special and occupies  the  leftmost vertex of the graph  (cf. Fig.~\ref{fig:graph_struct}). The action of  $H^+$ then corresponds to moving from left to right in the graph. This implies that the FSA recurrence closes after $N{+}1$ steps once forward propagation reaches the opposite edge of the graph, $ \ket{\mathbb{Z}'_2}$.

For a generic partial cube, the raising operator $H^+$ can be obtained directly from the graph. Indeed, once the starting vertex $j$ is chosen, all other vertices can be assigned a ``distance" $d_i{=}{\rm dist }(i,j)$ corresponding to the shortest path length from $i$ to $j$.
For partial cubes, the graph is bipartite and the distance is simply the Hamming distance. Because of this, there are no edges between states with the same index $d$. The graph can be turned into a directed graph by making all edges go towards the vertex with the higher index $d$.
The adjacency matrix of this graph is then the FSA raising operator $H^+$.

Now, the key property that enabled the FSA recurrence in the single hypercube case, Eq.~(\ref{Eq:backward-scatter}), continues to hold in the two hypercube model, but it is only satisfied approximately in the PXP model. 
More specifically, if one starts from the N\'eel state, Eq.~(\ref{Eq:backward-scatter}) is exact for $j=1,2$, but at the third step of the recurrence this property does not hold any more in the PXP model. Nevertheless, we can still enforce the FSA recurrence as defined in Eq.~(\ref{Eq:lanczosFSA}) and keep track of the incurred error.  The resulting vectors $\ket{v_j} \propto (H^+)^j \ket{v_0}$ obtained from the FSA recurrence, Eq.~(\ref{Eq:lanczosFSA}), starting from $\ket{v_0} = \ket{\mathbb{Z}_2}$, form an orthonormal subspace because each state belongs to a different Hamming distance sector and the recurrence closes after $N{+}1$ steps.  Diagonalising the tridiagonal matrix of size $(N{+}1){\times} (N{+}1)$ with $\beta_j$ determined either directly from Eq.~(\ref{Eq:lanczosFSA}) or via linear recurrence method explained in Ref.~\onlinecite{TurnerPRB}, one can obtain a set of approximate eigenenergies and eigenvectors. These eigenpairs turn out to be precisely the ones corresponding to $N{+}1$ scarred eigenstates in the PXP model and their analogues in the model of two hypercubes joined at a vertex. In Ref.~\onlinecite{Turner2017}  it was demonstrated that the eigenenergies agree within a few percent with exact diagonalization data for the largest available system of $N{=}32$  spins in the PXP model. Similarly, the expectation values of local observables and the entanglement entropy of scarred eigenstates are well-captured within the FSA~\cite{TurnerPRB}.

\subsection{FSA and the emergent su(2) algebra}

The FSA for the PXP model described above provides a foundation for understanding many-body scarred eigenstates as forming an approximate representation of a weakly ``broken" $\textrm{su(2)}$ Lie algebra~\cite{Choi2018,Bull2020}. The consequence of this algebra is an approximately decoupled  su(2) subspace in which revivals primarily take place, while the wavefunction amplitude slowly leaks into the thermal bulk (orthogonal subspace). The commutator of FSA operators $H^\pm$ allows us to define $H^z \equiv \frac{1}{2} [H^+, H^-]$. The set  $\{ H^+, H^-, H^z\}$ then forms a ``broken" su(2) Lie algebra~\cite{Bull2020} if they satisfy:
\begin{eqnarray}\label{eq:su2}
[H^z, H^{\pm}] = \pm H^{\pm} + \delta^{\pm},
\end{eqnarray}
provided that the operator norm of the correction term $|| \delta^\pm|| {\ll} 1$. Intuitively, this means that, up to a small error $|| \delta^\pm||$, we can view $H^\pm$ as the spin raising and lowering operators. Since the PXP Hamiltonian in Eq.~(\ref{eq:p_cube}) is a sum of the raising and lowering operators, $H_\mathrm{PXP} {=} H^+ + H^-$, it plays the role of a magnetic field in $x$-direction, hence the emergent spin will undergo precession when initialised in the $z$-polarised state $\ket{\mathbb{Z}_2}$.

Once a suitable raising operator and its corresponding FSA have been identified, there are two competing factors which influence the wave-function revivals. First, the FSA couplings $\beta_j$ must resemble those of an exact su(2) algebra in Eq.~(\ref{eq:betajmain}), which in turn dictates that the eigenvalues of the Hamiltonian, projected to the FSA subspace, must be approximately equidistant. In the two-hypercube model in Eq.~(\ref{eq:2hc}), the FSA is exactly disconnected from the rest of the Hilbert space but the couplings do not exactly match the su(2) values, hence the revivals are not perfect.
Second, even if couplings in the projected subspace are equivalent to an exact su(2) subspace, revivals may still decay if the wave function amplitude rapidly leaks into the orthogonal thermalising subspace. The quality of the FSA subspace can be characterised by the subspace variance:
\begin{eqnarray}\label{eq:subvar}
\sigma_E = \mathrm{Tr} \left[ P_{\textrm{FSA}}(H^2) - ( P_{\textrm{FSA}} (H ) )^2 \right],
\end{eqnarray}
where $P_\mathrm{FSA}$ is the projector to the FSA subspace.
In Refs.~\cite{Khemani2018,Choi2018,Bull2020} it was shown that the variance of the PXP scarred subspace can be suppressed by many orders of magnitude if certain perturbations are added to the PXP model, resulting in nearly perfect revivals over long times~\cite{Choi2018}.

\begin{figure}[tb]
\centering
  \centering
  \includegraphics[width=0.95\linewidth]{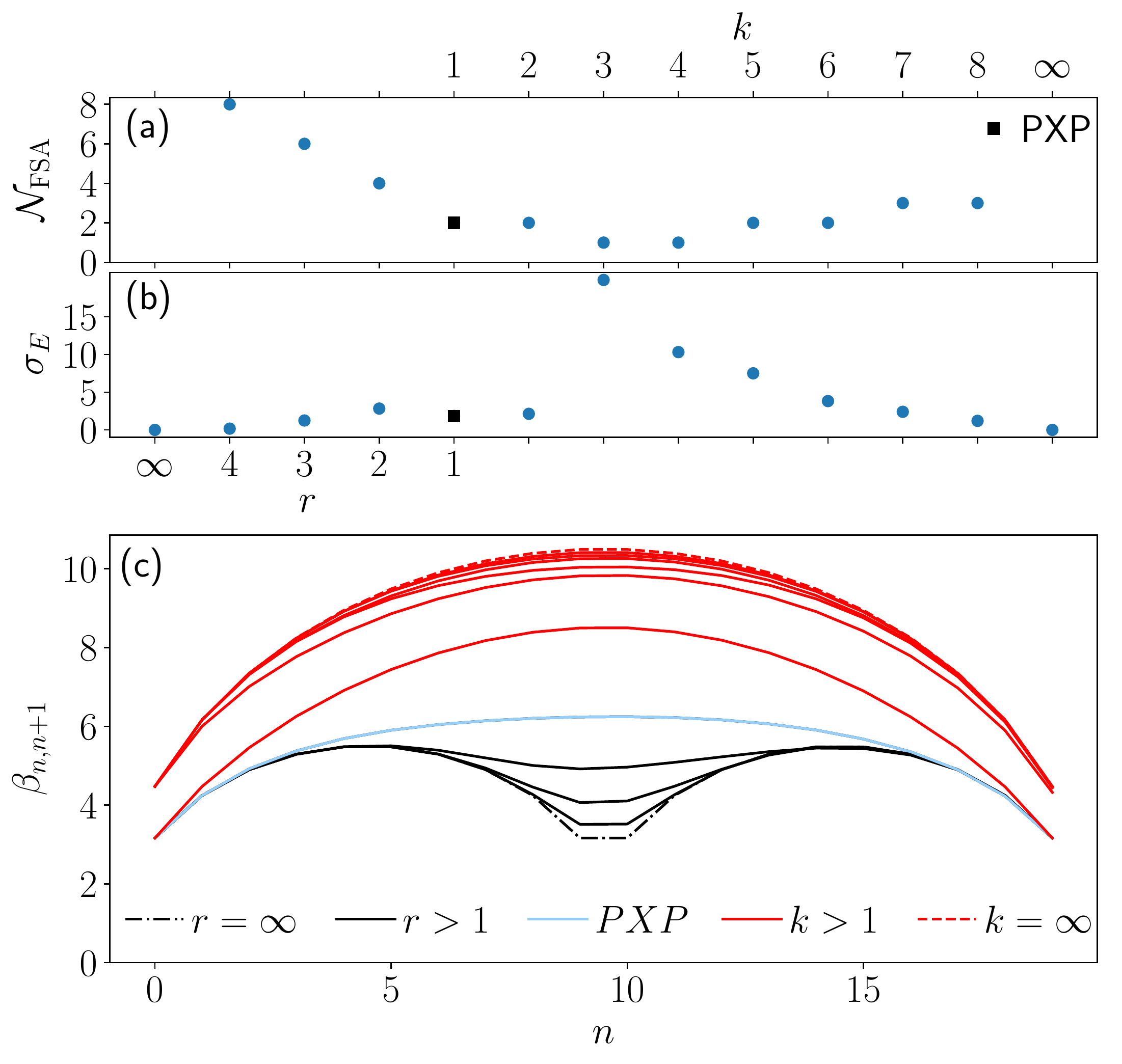}
  \caption{Number of exact FSA steps $\mathcal{N}_\mathrm{FSA}$, FSA subspace variance $\sigma_E$ and couplings $\beta$ for two families of models defined in   Eq.~(\ref{eq:2cubeinterp}) and Eq.~(\ref{eq:kmodels}) in the main text. These models interpolate between the two-hypercube model ($r=\infty$) and the free spin-1/2 model ($k=\infty$). The subspace variance remains relatively low for sufficiently strong constraint ($k{<}3$). This threshold value of $k$ coincides with the first step of the FSA changing. 
}
  \label{fig:FSA_all}
\end{figure}

\subsection{Effect of bridges on the FSA}

The FSA couplings $\beta_j$, subspace variance $\sigma_E$, and the number of exact FSA steps $\mathcal{N}_\mathrm{FSA}$ (i.e, steps where applying $H^-H^+$ leads back to the same state up to a multiplicative factor) are evaluated in Fig.~\ref{fig:FSA_all} for two families of constrained models defined by Eq.~(\ref{eq:2cubeinterp}) and Eq.~(\ref{eq:kmodels}) in the main text. The subspace variance can be seen to negatively correlate with the number of exact FSA steps, with a much higher value when only one step is exact. The number of exact steps can be well understood by the proximity to the two hypercubes or to the free spin-1/2 model, with the notable exception of $k{=}2$.

Indeed, for large values of $r$ the system is strongly constrained and the only difference with the two-hypercube model is the presence of states with a low density of excitations on both sublattices. These states are located near the shared vertex, which means that most FSA steps will be similar to the two-hypercube model and therefore exact. As $r$ is decreased, it is possible to add more excitations to one sublattice while still having a few excitations in the other. Consequently, the new states appearing in the system are increasingly closer to the N\'eel state. This means the number of FSA steps which are similar to the ones in the two hypercube model (and thus exact) decreases. Thus,  the subspace variance also increases, as can be seen in Fig.~\ref{fig:FSA_all}. This trend continues until $k=3$, where it is possible to add excitations directly to the N\'eel state and so even the first step of the FSA is unlike in the two cube model. This change in the first step can be clearly seen in the FSA coupling in Fig.~\ref{fig:FSA_all} (c). However, as it is now possible to remove any present excitation and to add any excitation on unexcited sites from the N\'eel state, the first FSA step is as in the free spin-1/2 model and so it is still exact.  
As $k$ is increased, it is possible to add more excitations on a sublattice even when the other is fully occupied. As a consequence, the number of FSA steps similar to the ones in the free spin-1/2 model increase and the subspace variance decreases.

One case that needs to be addressed separately is $k{=}2$. Indeed, as excitations can already be added in the second Hamming layer, we expect the corresponding FSA step to be inexact as it differs from the two-hypercube one. However, in practice we find that it is still exact. This supports the physical relevance of the 2HG subgraph, as it is only possible to branch out of this substructure at the third FSA step.

\section{System size and quantum dimension}\label{sec:rec}

In this Appendix, we derive quantum dimensions for the various families of models studied in the main text. To take the constraint into account, we make use of the transfer matrix $M$. Given a decomposition of states into classes, this matrix encodes how the class sizes for $N$ depend on the class sizes for $N{-}1$. The total number of states simply correspond to the sum of the class sizes. This subdivision of states allows one to make the effect of the constraint explicit. 

Let us take the PXP model as an example. We need to know how to ``glue'' a new site to one end of the chain. As the constraint forbids neighbouring excitations, we need to know the value of the leftmost site in the chain. Hence we will have two classes: states that end with an unexcited site and states that end with an excited site. The transfer matrix $M$ will thus be $2 \times 2$. The only forbidden process is adding an excited state next to an already excited, and the resulting transfer matrix is
\begin{equation}
M=\begin{pmatrix} 1 & 1 \\ 1 & 0 \end{pmatrix}.
\end{equation}
The Hilbert space dimension  $\mathcal{D}_N$ can be simply computed as $\rm{Tr}\left[M^N B\right]$ for $N\geq 1$, where $B$ is a matrix implementing the boundary conditions such that
\begin{eqnarray}
\notag B{=}\id \;\;\; \mathrm{for}\;\;\; \mathrm{PBC}, \;\;\; \mathrm{and} \;\;\;
B{=}\begin{pmatrix}1 \\ 0 \\ \ldots \\ 0\end{pmatrix}\begin{pmatrix}1 \ldots & 1\end{pmatrix} \;\;\; \mathrm{for}\; \mathrm{OBC}.\\
\end{eqnarray}
Indeed, inserting the identity lets the trace connect the rightmost and leftmost ends of the chain. For OBC, the column vector with a single non-zero entry represents the only state allowed \emph{past} the right end of the chain, which is an infinite trailing chain of unoccupied sites. The row vector has ones in all entries, as all possible configurations are allowed at the left end of the chain.

We can then directly derive the recursion relation for both OBC and PBC by computing the ordinary generating function
\begin{equation}
\begin{aligned}
    F(z)&=\sum_{N=0}^\infty z^N \mathcal{D}_{N} 
    =\sum_{N=1}^\infty z^N{\rm Tr}\left[M^N B\right] \\
    &={\rm Tr}\left[\left(\mathbf{1}-zM\right)^{-1} B\right]-{\rm Tr}\left[ B\right].
    \end{aligned}
\end{equation}
The recursion relation for $\mathcal{D}_{N}$ can be read off the denominator of the rational form of $F$.
For the PXP model we find that 
\begin{equation}
F_\mathrm{PBC}(z)=\frac{z(1+2z)}{1-z-z^2}, \quad F_\mathrm{OPBC}(z)=\frac{z(2+z)}{1-z-z^2}.
\end{equation}
It ensues that for both boundary conditions the recursion relation for the Hilbert space dimension of the PXP model is 
\begin{equation}
\mathcal{D}_N=\mathcal{D}_{N-1}+\mathcal{D}_{N-2},
\end{equation}
which is the well known Fibonacci recurrence.
The difference between boundary conditions stems from the initial conditions, as for PBC we have $\mathcal{D}_1{=}1$ and $\mathcal{D}_2{=}3$, while for OBC $\mathcal{D}_1{=}2$ and $\mathcal{D}_2{=}3$.

\begin{figure}[tb]
\centering
\includegraphics[width=\linewidth]{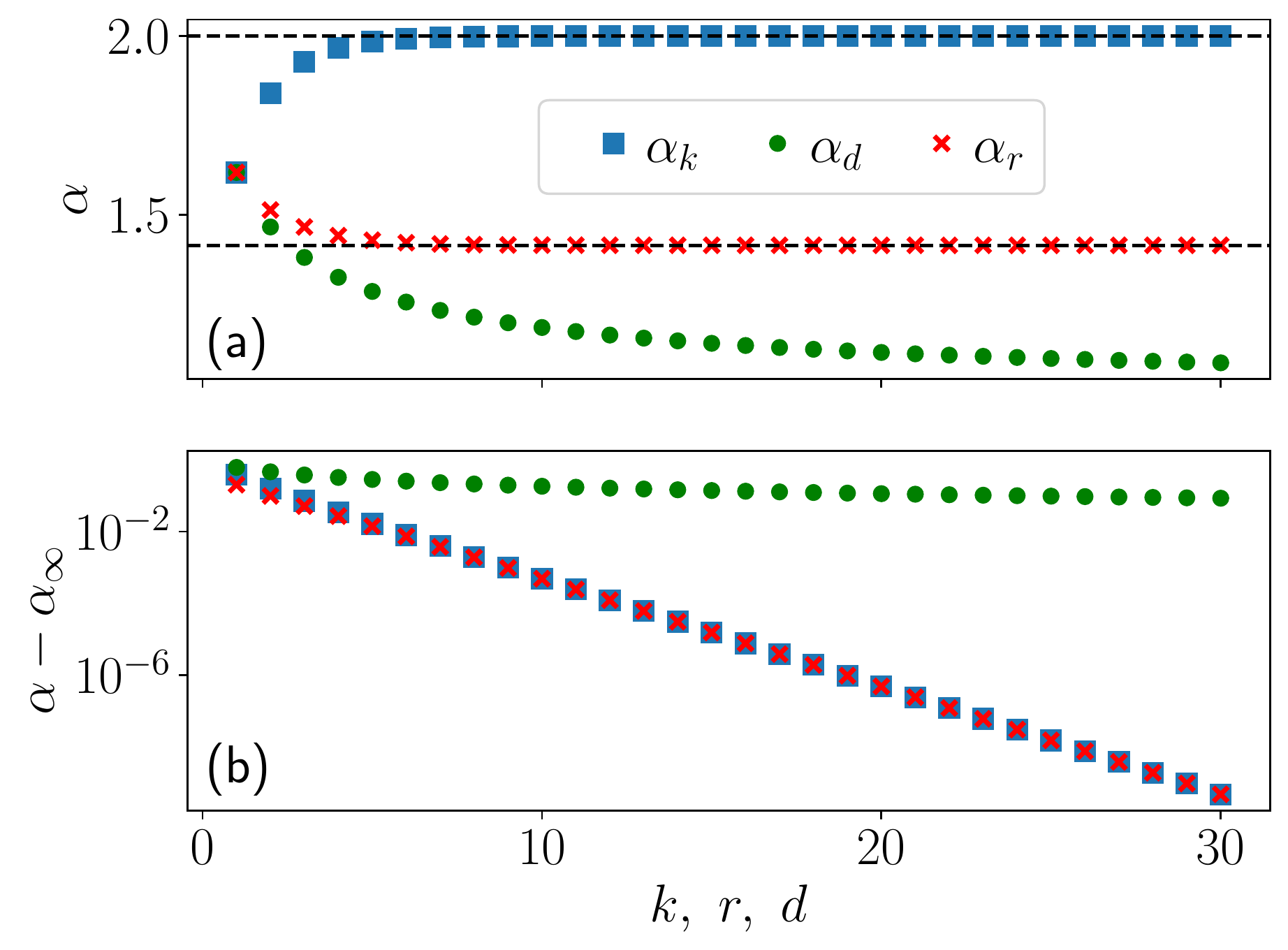}
\caption{Quantum dimension $\alpha$ for various constrained models studied in the paper. The dashed black lines correspond to $\alpha_{k=\infty}=2$, and $\alpha_{r=\infty}=\sqrt{2}$. For the $(k,k{+}1)$ and the models interpolating between PXP and the two hypercubes we see the same exponential convergence towards the asymptotic value.}
\label{fig:QM_dim}
\end{figure}

The quantum dimension 
\begin{eqnarray}\label{eq:quantumdim}
\alpha=\lim_{N \to \infty} \frac{\mathcal{D}_N}{\mathcal{D}_{N-1}}
\end{eqnarray}
can be found from the recursion relation by replacing $\mathcal{D}_N$ by $\alpha^N$ and finding the largest root.
For the PXP model we arrive at the equation $\alpha^2-\alpha-1=0$, which has the Golden Ratio $\phi{=}(1+\sqrt{5})/2$ as its largest-magnitude solution. Note that the quantum dimension can also be found as the largest-magnitude eigenvalue of the transfer matrix $M$. Below we use this transfer matrix and generating function formalism to derive the recursion relations and quantum dimensions for the various models investigated in this paper, whose summary is presented in Fig.~\ref{fig:QM_dim}.

\subsection{Models interpolating between the two hypercubes and PXP model}

For models defined in Eq.~(\ref{eq:2cubeinterp}), we need to take into account the $2r-1$ leftmost sites in order to glue a new one as this is the range of the constraint. However, as a single excitation is enough to saturate the constraint, we do not have to monitor the $2^{2r-1}$ combinations but only to track the position of the leftmost excitation. Because of this, the transfer matrix $M_r$ has size $2r \times 2r$ and the matrix elements are given by
\begin{eqnarray}
M_{i,j}=\delta_{i,1}\delta_{j,1}{+}\delta_{i,1}\delta_{j,2r}{+}\delta_{i,j+1}{+}\delta_{i,1}\sum_{l=1}^{\lfloor r/2 \rfloor}\delta_{j,2l{+}1},
\end{eqnarray}
giving
\begin{equation}\label{eq:M_r}
M_{r=2}=\begin{pmatrix} 1 & 0 & 0 & 1 \\ 1 & 0 & 1& 0 \\ 0 & 1 & 0 & 0 \\ 0 & 0 & 1 & 0\end{pmatrix}, \ M_{r=3}=\begin{pmatrix} 1 & 0 & 0 & 0 & 0 & 1 \\ 1 & 0 & 1 & 0 & 1 & 0 \\ 0 & 1 & 0 & 0 & 0 & 0 \\ 0 & 0 & 1 & 0 & 0 & 0 \\ 0 & 0 & 0 & 1 & 0 & 0 \\ 0 & 0 & 0 & 0 & 1 & 0\end{pmatrix},
\end{equation}
for $r{=}2$ and $r{=}3$.
For both boundary conditions, by using \texttt{Mathematica} to compute the denominator of the rational form of the generating function up to $r{=}80$, we found it to be $1-z-z^2-\sum_{j=3}^{2r}(-1)^j z^j$. Hence the recursion relation is
\begin{equation}
    \mathcal{D}_{N,r}=\mathcal{D}_{N-1,r}+\mathcal{D}_{N-2,r}+\sum_{j=3}^{2r}(-1)^j \mathcal{D}_{N-j,r}.
\end{equation}
For the initial conditions  $N\leq 2r$, for PBC we have $\mathcal{D}_{N,r}=1$ if $N$ is odd and $\mathcal{D}_{N,r}=2^{N/2+1}-1$ if $N$ is even, while for OBC $D_{N,r}=2^m+2^{N-m}-1$, where $m=\lceil N/2 \rceil$.

From the recursion relation, the quantum dimension is the largest root of the equation 
\begin{equation}
\alpha_r^{2r}-\alpha_r^{2r-1}-\alpha_r^{2r-2}-\sum_{j=3}^{2r}(-1)^j \alpha_r^{2r-j}=0.
\end{equation}
As $r{\rightarrow}\infty$, it must hold that $\alpha_r{\rightarrow} \sqrt{2}$ as the corresponding model is the two-hypercube one with $\mathcal{D}_{N,r=\infty}=2^{\frac{N}{2}+1}+1$, as we indeed see in Fig.~\ref{fig:QM_dim}. 

\subsection{Larger blockade radius}

When the Rydberg blockade radius is extended to the neighbouring $d$ sites, Eq.~(\ref{eq:ppxpp}), it is fairly easy to derive the transfer matrix. Indeed, we only need to know if there is an excitation in the leftmost $d$ sites and, if yes, its position. Hence the transfer matrix now has size $d+1 \times d+1$ and the matrix elements are given by
\begin{eqnarray}
M_{i,j}=\delta_{i,1}\delta_{j,1}+\delta_{i,d+1}\delta_{j,1}+\delta_{i+1,j}.
\end{eqnarray}
Specifically,
\begin{equation}\label{eq:M_d}
M_{d=2}=\begin{pmatrix} 1 & 1 & 0 \\ 0 & 0 & 1 \\ 1 & 0 & 0\end{pmatrix}, \ M_{d=3}=\begin{pmatrix} 1 & 1 & 0 & 0 \\ 
0 & 0 & 1 & 0 \\ 0 & 0 & 0 & 1 \\ 1 & 0 & 0 & 0\end{pmatrix}~.
\end{equation}

The denominator of the rational generating function is then $1-z-z^{d+1}$ for both boundary conditions. This was tested using \texttt{Mathematica} for $d$ up to 100. The recursion relation is then
\begin{equation}
    \mathcal{D}_{N,d}=\mathcal{D}_{N-1,d}+\mathcal{D}_{N-d-1,d}.
\end{equation}
The initial conditions $N\leq d+1$ are simple to derive, as for OBC $\mathcal{D}_{N,d}=N+1$ while for PBC it is $\mathcal{D}_{N,d}=1$ for $N<d+1$ and $\mathcal{D}_{N,d}=N+1$ for $N=d+1$.

Alternatively, there is a simple argument for the OBC recursion relation. The term $\mathcal{D}_{N-1,d}$ trivially counts all configurations with a leftmost non-excited site. On the other hand, all models with a leftmost excited state must have the $d$ next sites unexcited, and so the number of configuration of the $N-1-d$ remaining sites is equal to  $\mathcal{D}_{N-d-1,d}$. 

The recursion relation implies that the quantum dimension is the largest root of the equation
\begin{equation}
\alpha_d^{d+1}-\alpha_d^d-1=0
\end{equation}
As $d\rightarrow\infty$, $\alpha_d\rightarrow 1$ as the corresponding model can only hold a single excitation at a time and so $\mathcal{D}_{N,d=\infty}=N+1$.

\subsection{$(k, k{+}1)$ models}\label{sec:kkp_supp}

Finally, for $(k, k{+}1)$ models in Eq.~(\ref{eq:kmodels}), we need to keep track of the exact number and location of excitations in the $k$ leftmost sites. As a consequence, we need to have $2^k$ classes corresponding to the individual configurations. The resulting transfer matrix has size $2^k \times 2^k$ and its matrix elements are given by 
\begin{eqnarray}
M_{i,j} = \delta_{j,\lceil i/2 \rceil}+(1-\delta_{j,2^k})\delta_{j,2^{k-1}+\lceil i/2 \rceil}.
\end{eqnarray}
For example, for $k{=}1,2$ we have 
\begin{equation}\label{eq:M_k}
M_{k=1}=\begin{pmatrix} 1 & 1 \\ 1 & 0\end{pmatrix}, \ M_{k=2}=\begin{pmatrix} 1 & 0 & 1 & 0 \\ 
1 & 0 & 1 & 0 \\ 0 & 1 & 0 & 1 \\ 0 & 1 & 0 & 0\end{pmatrix}~.
\end{equation}
The numerator of the generating function is found to be $1-\sum_{j=1}^{k+1}z^j$ for both OBC and PBC boundary conditions.
We were able to compute it up to $k{=}9$ using \texttt{Mathematica}. 
As a consequence the recursion relation is
\begin{equation}\label{eq:rec}
    \mathcal{D}_{N,k}=\sum_{j=1}^{k+1}\mathcal{D}_{N-j,k}.
\end{equation}
The values of $\mathcal{D}_{N,k}$ for $N\leq k$ do change with the boundary conditions, as with OBC we have $\mathcal{D}_{N,k}=2^N$  while for PBC it is $\mathcal{D}_{N,k}=2^N-1$. For $N{=}k+1$ we have $\mathcal{D}_{N,k}=2^{k+1}-1$ in both cases.
As $k\rightarrow\infty$, $\alpha_k\rightarrow 2$ as the corresponding model is the free spin-1/2 model with $\mathcal{D}_{N,k=\infty}=2^N$. 

Eq.~(\ref{eq:rec}) implies that the quantum dimension $\alpha_k$ must satisfy
\begin{equation}
    \alpha_k^{k+1}-\sum_{j=0}^{k}\alpha_k^{k}=0.
\end{equation}

\section{Random bridges on two connected hypercubes}\label{sec:algo}

In order to test how sensitive our conclusions in the main text are to the details of the graph structure, we devised a protocol for random sampling of models that interpolate between two joined hypercubes and the full hypercube. As two joined hypercubes of dimension $N/2$ are contained in a hypercube of dimension $N$, the protocol works by adding back states from the full hypercube to the two smaller hypercubes.  In order to match the constraints in the considered models, at each step the Hamiltonian  can be written as Eq.~(\ref{eq:p_cube}) with constraints only on \emph{excitations} (meaning that it is always possible to remove excitations) and with translation symmetry conserved. The process also ensures that the graph remains unweighted, i.e., the matrix elements of the Hamiltonian are all equal.

To formalise the protocol, let us denote each basis state by a binary string $u{\in} B^N$, with $B{=}{0,1}$.
Consider two states $u$ and $v$;  we say that $u{\leq} v$ if $u_i {\le} v_i$ for $i{=}1$ through $N$.
This is strictly equivalent to saying that $u$ can be obtained by only removing excitations from $v$.
Then, in all these models, if $v{\in} G$ and $u{\leq} v$, then $u{\in} G$ as well. 
Because an excitation can always be removed, all these models are ``daisy cubes" \cite{Klavzar2019}:
\begin{dfn}
A daisy cube is defined by a $N$ dimensional hypercube graph $\HG{2}{N}$ and a set of states $X$, such that all elements of $X$ are in $\HG{2}{N}$.
Then the set of vertices in the corresponding daisy cube is defined as $V(\HG{2}{N}(X))=\{v\in \HG{2}{N} | \exists x\in X\ \text{s.t}\ v\leq x\}$.
The graph $\HG{2}{N}(X)$ is the subgraph of $\HG{2}{N}$ induced by $V(\HG{2}{N}(X))$.
Equivalently, there is an edge between two states in $\HG{2}{N}(X)$ if their strings differ by a single element.
\end{dfn}

This formulation also implies that different sets $X$ can correspond to the same graph.
In particular, if $x,y\in X$ and $y\leq x$, then $\HG{2}{N}(X)=\HG{2}{N}(X \setminus \{y \})$.
However it is clear that there exists a unique set $\hat{X}$ of maximal vertices such that it has the minimum cardinality of all sets representing the same daisy cube.
For the PXP model in Eq.~(\ref{eq:p_cube}) with $N{=}6$ for example, the maximal vertices set is
\begin{equation*}
\hat{X}_{\text{PXP},N=6}=\{101010, \ 010101, \ 100100, \ 010010, \ 001001\}.
\end{equation*}
The graph which has $H$ as its adjacency matrix is the daisy cube $\HG{2}{6}(\hat{X}_{\text{PXP},N=6})$.
Importantly, daisy cubes are all partial cubes~\cite{Klavzar2019}.

We will use the binary string and daisy cubes notations to define the sampling algorithm.
The translation operator $T$ acts on the strings as $(Tu)_i{=}u_{i-1}$, with periodic boundary conditions such that $u_N{=}u_0$.
Because of translation symmetry, the two stitched hypercubes correspond to the daisy cube $\HG{2}{N}(X), X=\{\mathbb{Z}_2, T\mathbb{Z}_2\}$, where $\mathbb{Z}_2$ is the N\'eel state $1010 \ldots 10$.
The hypercubes are added in increasing dimension from $k{=}2$ to $k{=}N$, and after each addition the revivals from the N\' eel state are computed.
The interpolation parameter $\lambda$ is also computed at each step, as defined in Eq.~(\ref{eq:bridgedensity}) in the main text.
The exact algorithm is :
\begin{figure}[H]
\begin{algorithm}[H]
  \caption{Random bridges on 2 hypercubes}
  \label{Algo:bridges}
   \begin{algorithmic}[0]
   \State Set $X=\{\mathbf{Z}_2,T\mathbf{Z}_2 \}$
   \State Set the graph as $G=\HG{2}{N}(X)$
   \State Set $k=2$
   \While {$k\leq N$}
   \State Pick $u\in \HG{2}{N}$ such that $\sum_{i=1}^N u_i {=}k$, $u\nleq \mathbf{Z}_2$ and $u\nleq T\mathbf{Z}_2$  
   \If {$u\in G$} 
      \State $k=k+1$
  \Else
       \For {$r=0$ to $N-1$}
       \State $X=X \cup \{T^r u\}$
       \EndFor
       \State Update the graph as $G=\HG{2}{N}(X)$
       \State Get the Hamiltonian as $H=\text{Adj}(G)$
       \State Compute $\lambda (G)$
       \State Compute the revivals from the N\'eel state

   \EndIf
  \EndWhile

   \end{algorithmic}
\end{algorithm}
\end{figure}

\section{The effect of low-density bridges on the two connected hypercubes}\label{sec:2-HC}

In the main text we showed that adding bridges to the two-hypercube model at first leads to an increase of revival fidelity at small $\lambda$, before a decrease at larger $\lambda$ values. In this section we examine in detail the former regime and show that it can be understood as a tuning of the symmetric and anti-symmetric sectors of the two-hypercube model.

\begin{figure}[htb]
\centering
\includegraphics[width=\linewidth]{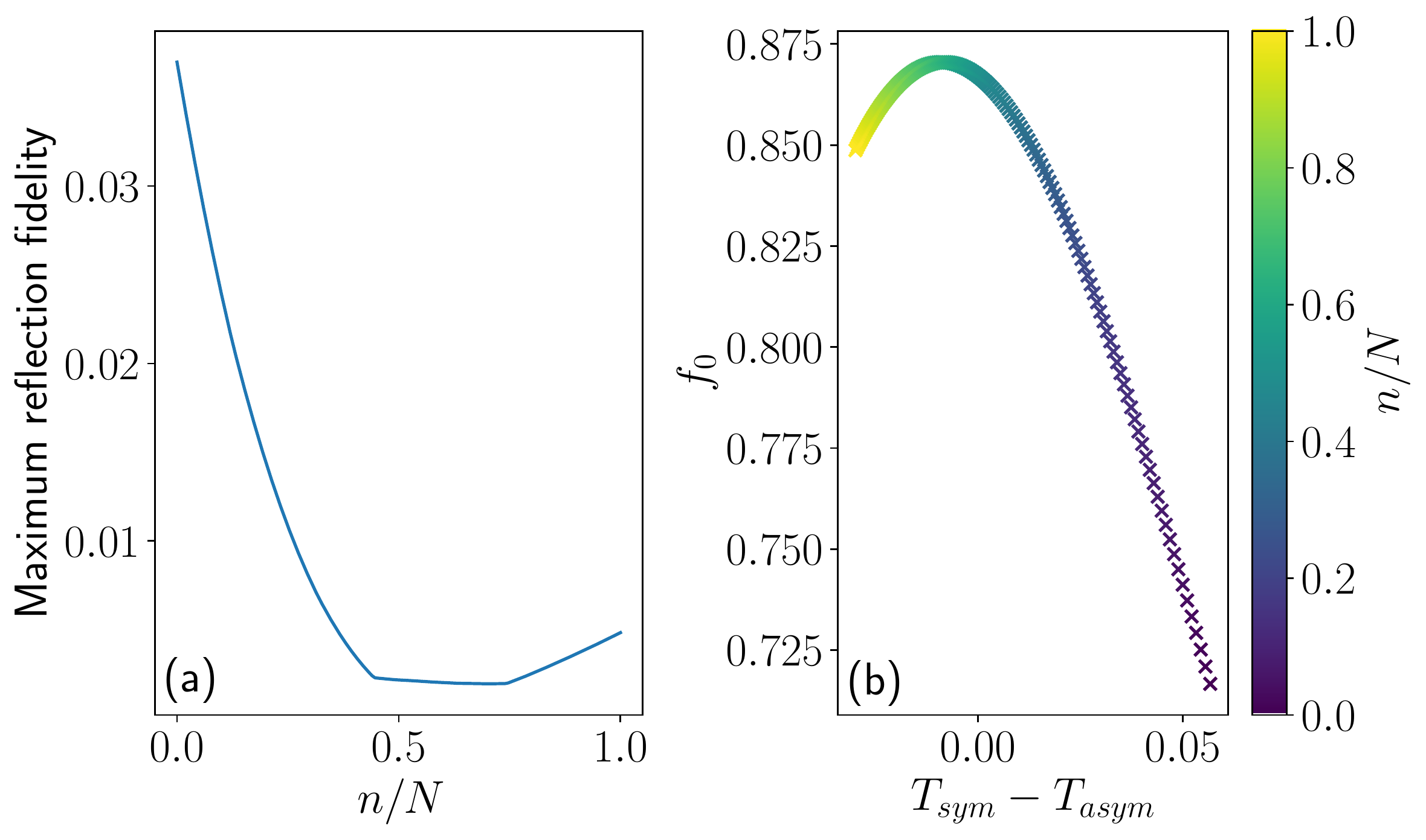}
\caption{Fidelity of revivals and reflection for two linked hypercubes of dimension $N{=}300$ as two-dimensional bridges are added. (a) Maximum fidelity of the reflection peak. (b) Fidelity of revivals with respect to the period difference between the symmetric and antisymmetric sectors.}
\label{fig:2HC_2D_fid}
\end{figure}

We focus on the first steps of the algorithm described in Appendix~\ref{sec:algo}, when the number of excitations of the vertices added is equal to 2. We will refer to these as bridges of dimension 2.
As all states with two excitations on the same sublattice are already in the two hypercubes, all vertices added will have one excitation on each sublattice. 
These states are all located on the Hamming layer $N{+}1$, the  same Hamming layer as the shared vertex that contains no excitations.
Intuitively, in the many-body picture one can think these new states as helping ``spread" the support of the wave function on several states instead of just ``funneling" the wave function onto a single state, $\ket{000\ldots 000}$.
This means that even if the wave function still reflects from  this vertex, the reflected part of the wave function will have smaller magnitude, as there is no longer a high concentration of the wave function on this vertex -- see Fig.~\ref{fig:2HC_2D_fid} (a).
However, as more bridges of dimension two are added, the degree of the vertices in the Hamming layers $N$ and $N+2$ increases, and this could lead to reflection in this layer.
In order to completely get rid of reflection one would need a smooth coupling profile, which can be achieved by also adding vertices in other Hamming layers, like in the PXP model.
This would help get rid of reflection but at the price of having an inexact FSA.
This is in line with what we observe as random bridges are added: no reflection but a leakage outside the FSA subspace.

For two-dimensional bridges, enforcing translation symmetry has the effect of modifying the edges between the Hamming layers $N$, $N{+}1$ and $N{+}2$ in an isotropic fashion.
As a consequence, the FSA remains exact but the middle couplings are changed. 
Normally, both of these couplings are equal to $\beta_\mathrm{middle}{=}\sqrt{N}$, however  adding  $V$ vertices changes the coupling to $\beta_\mathrm{middle}{=}\sqrt{N+V/N}$.
In order to simplify computations, let us assume that $N$ is even.
Then there are $N/2$ bridges that can be added such that they are not equivalent under translation, and adding any of these implies adding $2N$ vertices.
This means that the middle FSA couplings can take values $\beta_\mathrm{middle}{=}\sqrt{N+n}$, $n{=}0,2,4,\ldots,N$, hence $\sqrt{N}{\leq} \beta_\mathrm{middle} {\leq} \sqrt{2N}$.
Furthermore, it means that details of the bridges do not matter, but only their number.
So all random processes will be identical if only hypercubes of dimension 2 are added.
The results of this process for two joined hypercubes of different sizes can be seen in Fig.~\ref{fig:2HC_2D_fid} (b), where the color indicates the density of these bridges. 

The results in Fig.~\ref{fig:2HC_2D_fid} can be understood in terms of the the two symmetry sectors mentioned in Sec.~\ref{sec:twocubes}.  Indeed, the bridges considered here only affect the symmetric sector, changing its coupling between the last two states from $\sqrt{2N}$ to $\sqrt{2(N+n)}$. This reduces the revival period of this sector, making it closer to the period of the anti-symmetric sector until it overshoots and makes them further apart.
This can be seen in Fig.~\ref{fig:2HC_2D_fid}, where the correlation between the density of bridges $n/N$, the revival fidelity, and the period difference between the sectors is apparent.
This also makes the reflection peak much smaller, as the two sectors almost exactly cancel out at $T\approx \pi$.
The only thing preventing the reflection to be exactly 0 is the difference of revival amplitude between them.

\section{More than two hypercubes in a star configuration}\label{sec:n-cubes}

\begin{figure}[t]
\centering
\includegraphics[width=\linewidth]{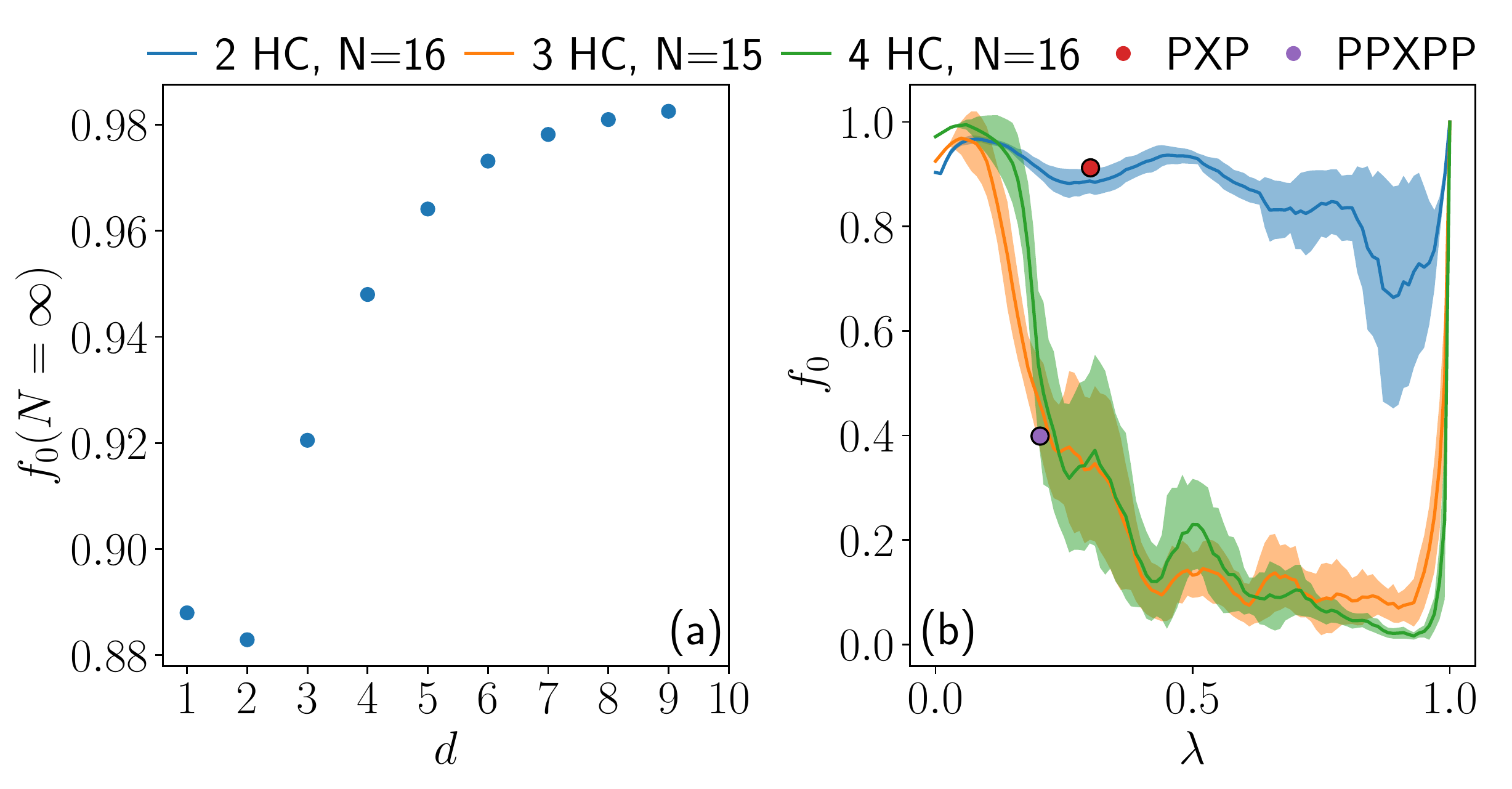}
\caption{(a) Revivals for $d+1$ joint hypercubes in the symmetric sector in the thermodynamic limit from finite size scaling. (b) Evolution of the first revival peak $f_0$  upon the addition of bridges in the symmetric sector of multiple hypercubes arranged in a star pattern. The case with two hypercubes (2HC or $d{=}1$) is seen to be radically different from three hypercubes (3HC, $d{=}2$) and four hypercubes (4HC, $d{=}3$).
}
\label{fig:HC_k0}
\end{figure}

In the main text we mentioned that models with a Rydberg blockade radius $d$, Eq.~(\ref{eq:ppxpp}), naturally realise $d+1$ hypercubes joined at a single vertex, forming a star-like pattern. Like in the two connected hypercubes, the Hamiltonian of $d+1$ hypercubes arranged in this manner can be reduced to $d+1$ tight-binding chains linked at a single site. 
If we label these chains $L_j$, then we can reorganise the Hilbert space into $d+1$ symmetry sectors $S_{k}=\sum_{j=0}^{d} e^{2\pi i k j/(d+1)}L_{j}$, with $k=0,1,\ldots, d$. 
The only difference between the symmetry sectors is the shared vertex. 
The totally symmetric sector has $N{+}1$ states and its Hamiltonian is the one from Eq.~(\ref{eq:HCTB}) except that the last term of the sum is multiplied by a factor of $\sqrt{d+1}$. In the non-symmetric sectors, the contribution of the $d+1$ chains cancel at the middle vertex, so  these sectors are all identical. 
As the symmetric sector is the only one affected by the blockade range we focus on it.
Finite size scaling indicates that in the thermodynamic limit revivals are present in it for all values of $d$, even though the first revival fidelity is non-monotonic in $d$ -- see Fig.~\ref{fig:HC_k0}.

We also studied the effect of bridges in the symmetric sector. Doing this for two, three and four hypercubes in a star pattern shows a clear difference between two hypercubes and other cases -- see Fig.~\ref{fig:HC_k0} (b).  This difference is attributed to the change in the reviving subspace.
As we discussed in Sec.~\ref{sec:alt}, for $d$ joined hypercubes the revivals occur because of state transfer between the state $|\mathbb{Z}_d\rangle$ in Eq.~(\ref{eq:zd}) and its translations. Meanwhile, for the full hypercube the state transfer is generally different, recall Eq.~(\ref{eq:startransfer}).
Thus, the subspace in which the revivals happen changes drastically during the interpolation.
For a large value of lambda ($\lambda\approx 0.5$) the dynamics is no longer restricted to an almost closed subspace but  can spread into the Hilbert space.
In contrast, for $d{=}1$ the reviving subspace stays relatively unchanged. 
Indeed, the new states added do not change the dimension of the reviving subspace (the FSA), but only modify the couplings and the subspace variance.
While this naturally affects the revivals, it does it in a much less drastic way than changing the subspace altogether. 

\bibliography{references}

\end{document}